\documentclass{LMCS}
\makeatletter
\let\@titlecomment=\@empty
\makeatother
\usepackage{calc,gastex,hyperref} 

\newcommand{\dOi}[1]{\quad\href{http://dx.doi.org/#1}{doi:#1}}

\newcommand{\para}[1]{\medskip\noindent{\bfseries #1.}}

\newcommand{\natN}{{\mathbb N}}
\renewcommand{\vec}[1]{\overline{#1}}
\newcommand{\tuple}[1]{\langle #1 \rangle}

\newcommand{\op}[1]{\operatorname{#1}}

\newcommand{\rank}{\operatorname{rank}}
\newcommand{\lab}{\operatorname{lab}}
\newcommand{\edg}{\operatorname{edg}}
\newcommand{\rt}{\operatorname{root}}

\newcommand{\family}[1]{\mbox{\sf #1}}
\newcommand{\logic}[1]{\mbox{#1}}

\newcommand{\negg}{{\sim}}  % additional brackets to get rid 
\newcommand{\cA}{{\mathcal A}}       % standard automaton
\newcommand{\DTWA}[1]{\family{DW$^{#1}$A}}
\newcommand{\NTWA}[1]{\family{NW$^{#1}$A}}
\newcommand{\ATWA}[1]{\family{AW$^{#1}$A}}
\newcommand{\DPTWA}[1]{\family{DPW$^{#1}$A}}
\newcommand{\NPTWA}[1]{\family{NPW$^{#1}$A}}
\newcommand{\APTWA}[1]{\family{APW$^{#1}$A}}

\newcommand{\tc}{\operatorname{\sf tc}}
\newcommand{\REGT}{\family{REG}}
\newcommand{\FO}{\family{FO}}
\newcommand{\FODTC}[1]{\family{FO+DTC$^{#1}$}}
\newcommand{\FOTC}[1]{\family{FO+TC$^{#1}$}}
\newcommand{\FOposD}[1]{\family{FO+posDTC$^{#1}$}}
\newcommand{\FOpos}[1]{\family{FO+posTC$^{#1}$}}
\newcommand{\LFO}{\family{LFO}}
\newcommand{\MSO}{\family{MSO}}

\newcommand{\NLOG}{\family{NSPACE($\log n$)}}
\newcommand{\DLOG}{\family{DSPACE($\log n$)}}

\newenvironment{theorem}{\begin{thm}}{\end{thm}}
\newenvironment{lemma}{\begin{lem}}{\end{lem}}
\newenvironment{example}{\begin{exa}}{\end{exa}}
\newenvironment{corollary}{\begin{cor}}{\end{cor}}
\def\doi{3 (2:3) 2007}
\lmcsheading%
{\doi}
{1--27}
{}
{}
{Apr.~\phantom{0}9, 2006}
{Apr.~26, 2007}
{}

\begin{document}
\title[Nested Pebbles and Transitive Closure]
  {Automata with Nested Pebbles Capture\\
   First-Order Logic with Transitive Closure}
\author[J.~Engelfriet]{Joost Engelfriet}
\address{Leiden University, Institute of Advanced Computer Science, 
P.O.Box 9512, 2300 RA Leiden, The Netherlands}
\email{\{engelfri,hoogeboom\}@liacs.nl} 

\author[H.~J.~Hoogeboom]{Hendrik Jan Hoogeboom}
%\address{second author same address as the first author}

%% ===============================================
%% ===============================================
\begin{abstract}
String languages recognizable in (deterministic) log-space
are characterized either by two-way (deterministic) multi-head automata,
or following Immerman, by first-order logic with (deterministic) transitive closure.
Here we elaborate this result, and match the number of heads
to the arity of the transitive closure.
More precisely,
first-order logic with 
$k$-ary deterministic transitive closure
has the same power as 
deterministic automata walking on their input with $k$ heads,
additionally using a finite set of nested pebbles.
This result is valid for strings, ordered trees, 
and in general for families of graphs having 
a fixed automaton that can be used to traverse the nodes
of each of the graphs in the family.
Other
examples of such families are grids, toruses, and rectangular mazes.
For nondeterministic automata, the logic is restricted to
positive occurrences of transitive closure.

The special case of $k=1$ for trees, 
shows that single-head deterministic tree-walking automata 
with nested pebbles
are characterized by first-order logic with 
unary deterministic transitive closure.
This refines
our earlier result that placed these automata
between first-order and monadic second-order logic on trees.
\end{abstract}

\maketitle

%% ===============================================
%% ===============================================
\section{Introduction}
%% ===============================================

The complexity class \DLOG\ of string languages 
accepted in logarithmic space by deterministic Turing machines,
has two well-known distinct characterizations.
The first one,  
see e.g., \cite{Har72}
(or Corollary~3.5 of \cite{Iba71}),
is in terms of deterministic two-way automata with several heads 
working on the input tape
(and no additional storage).
Second, Immerman \cite{Imm87}
showed that these languages can
be specified using first-order logic with an additional 
deterministic transitive
closure operator -- it is one of the main results in the
field of descriptive complexity \cite{EbbFlu,Imm99}.
Similar characterizations 
of \NLOG\ 
hold for their nondeterministic counterparts
\cite{Imm87,Imm88,Sze}.

So we have a match between two distinct ways of specifying
languages
(two-way multi-head automata and first-order logic with
transitive closure)
that both have a natural parameter indicating the relative complexity
of the mechanism used.
For multi-head automata the parameter is the number of heads used
to scan the input; indeed,
two heads are more powerful than a single one,
as two heads can be used
to accept nonregular languages like $\{ a^nb^n\mid n \in \natN \}$,
whereas single-head two-way automata only accept regular languages
\cite{RabSco,She}.
In fact, it is known that in general $k+1$ heads are better
than $k$ even for a single-letter input alphabet \cite{Mon}.
For transitive closure logics, the parameter is 
the arity of
the transitive closure operators used:
if $\phi(\vec{x},\vec{y})$ denotes a formula that relates two
sequences $\vec{x},\vec{y}$ of $k$ variables each, then
$\phi^*(\vec{x},\vec{y})$ denotes the transitive (and reflexive)
closure of $\phi$ --
we will call this $k$-ary transitive closure,
and it is said to be deterministic if $\phi$ determines
$\vec{y}$ as a function of $\vec{x}$.
%=
Clearly, binary transitive closure is more powerful than unary:
$\{ a^nb^n \mid n\in \natN \}$ can easily be described in
first-order logic with binary deterministic transitive closure,
but first-order logic with unary transitive closure defines the 
regular languages \cite{BarMak,EbbFlu,Pott}.
It seems to be open whether ($k+1$)-ary transitive closure is
more powerful than $k$-ary transitive closure, see \cite{Gro}.

In \cite{BarMak}, Bargury and Makowsky set out to characterize
the formulas that capture the power of automata with $k$ heads, 
and they found a class of formulas, called $k$-regular,
with this property. Apart from first-order concepts, the formulas
only use $k$-ary transitive closure.
They show how $k$-head automata can be described by $k$-ary transitive
closure (both deterministically and nondeterministically)
but for the converse the $k$-ary regular formulas
only work in the nondeterministic case:
``the modification of the $k$-regular formulas needed to take
out the nondeterminism will spoil their elegant form, 
and we do not pursue this further'' \cite{BarMak}.

Here we set out from the other side. Starting with the full
set of deterministic 
$k$-ary transitive closure formulas we want to obtain an equivalent 
notion of deterministic automata.
Indeed, we succeed in doing this, 
i.e., we have an automata-theoretic characterization of
first-order logic with deterministic 
$k$-ary transitive closure,
but we pay a price.
The two-way automaton model we obtain has $k$ heads, 
as expected,
but is augmented with the possibility to put an arbitrary
finite number
of pebbles on its input tape, to mark positions for further use.
If these pebbles can be used at will it is 
folklore \cite{RitSpr,Pet} that
we obtain again \DLOG, a family too large for our purpose.
Instead we only allow pebbles that are used in a 
LIFO (or nested) fashion: 
all pebbles can be `seen' by the automaton as usual, 
but only the last one dropped can be picked up
\cite{GloHar,walking,MilSV,Via,EngMan,NevSchVia}.
On the other hand our pebbles are more flexible than the usual
ones: they can be `retrieved from a distance', i.e.,
a pebble can be picked up even when no head is scanning the
position of the pebble
(cf. the ``abstract markers'' of \cite{BluHew}).

In the nondeterministic case we have to restrict ourselves to
formulas with positive occurrences of the 
$k$-ary transitive closure operator, as we do not know whether
the class of languages accepted by nondeterministic $k$-head 
two-way automata with nested pebbles is closed under
complement.

In fact, our equivalence result (Theorem~\ref{main}) is stated 
and proved for ranked trees in general,
of which strings are a special case. 
Our automaton model (a tree-walking automaton) visits the
nodes of an input tree,  moving up and down along the
edges of the tree. The moves  (and state changes)
are determined by the state of
the automaton and the label of (and pebbles on) 
the nodes it visits;
additionally we assume that the children of each node are
consecutively numbered and that the automaton  
can distinguish this number.

In Section~\ref{logpeb} we translate logical formulas
into automata, following \cite{walking}
and additionally using the technique of Sipser \cite{Sip}
to deterministically search a computation space.
Section~\ref{peblog} considers the reverse:
translating automata into logical formulas.
As in \cite{BarMak} we adapt Kleene's construction
to obtain regular formulas from automata,
thus getting rid of the states of the automaton 
(Lemma~\ref{Phi}),
but we need to iterate that construction:
once for each nested pebble.

In Section~\ref{overview} we summarize the results of our paper
for single-head automata on trees, 
the context of our previous papers
\cite{walking,trips}
dealing with tree-walking automata and the quest of obtaining
simple tree automaton models that are inherently sequential 
(unlike the classic tree automata) and still capture the full power
of regular tree languages.

Finally, in Section~\ref{graphs}
we discuss how to extend our results to more general
graph-like structures,
such as unranked trees 
(important for XML \cite{MilSV,Via,XML,KlaSchSuc,Sch}),
cycles, grids (as in \cite{BarMak};
important for picture recognition \cite{BluHew,Ros,GiaRes,MatSchTho,Mat}),
toruses, and, for $k\ge 2$,
mazes \cite{BluKoz,Bud,Hof}.
To have a meaningful notion of graph-walking automaton we only
consider graphs with a natural locality condition: a node
cannot have two incident edges with the same label and the
same direction. 
Note that unranked
trees satisfy this condition when we view them as graphs with
`first child'
and `next sibling' edges in the usual way. 
Two-dimensional grids satisfy
it by distinguishing between horizontal and vertical edges.
For all such graphs, our result holds in one direction: from
automata to logical formulas. 
The other direction holds for all families of such graphs for
which there exists a single-head 
deterministic graph-walking automaton (with
nested pebbles) that can traverse each graph of the family,
visiting each node at least once.
This includes all families mentioned above (except mazes, for
which the automaton has two heads), and also, trivially, the
family of `ordered' graphs, i.e., all graphs in which the
successor relation of a total order is determined by edges
with a specific label.
Note that the existence of such an automaton is needed to
implement even simple logical formulas such as 
`all nodes have label $\sigma$'.

\bigskip
The main result of this paper, but only for trees and $k=1$
(cf. Section~\ref{overview}),
was first presented at a workshop in Dresden in March 1999,
but unfortunately did not make it into the proceedings \cite{Vog99}.
Muscholl, Samuelides, and Segoufin \cite{MuSaSe} 
have `reconstructed' our missing result,
independently obtaining the closure under complementation
of the tree languages accepted by
deterministic tree-walking automata with nested pebbles,
taking special care to minimize the number of pebbles needed.
The results of this paper were then presented at
STACS~2006 \cite{STACS}.

%% ===============================================
%% ===============================================
\section{Preliminaries}\label{sec:prel}
%% ===============================================

\para{Trees}
A \emph{ranked alphabet} is a finite set $\Sigma$ together with
a mapping $\rank: \Sigma \to \natN$.
Terms over $\Sigma$ are recursively defined:
if $\sigma \in \Sigma$ is of rank $n$, 
and $t_1,\dots,t_n$ are terms, 
then $\sigma(t_1,\dots,t_n)$ is a term.
In particular $\sigma$ is a term for each symbol $\sigma$ of rank 0.

As usual, terms are visualized as \emph{trees},
which are special labelled graphs;
$\sigma(t_1,\dots,t_n)$ as a tree which has a root
labelled by $\sigma$ and outgoing edges labelled by
$1,\dots,n$ leading to the roots of trees for
$t_1,\dots,t_n$. 
The roots of subtrees $t_1,\dots,t_n$ are said to have
child number $1,\dots,n$, respectively; by default the child number of the
root of the full tree equals $0$.
The set of all trees (terms) over ranked alphabet 
$\Sigma$ is denoted by $T_\Sigma$.

Strings over an alphabet $\Sigma$ can be seen as a special case:
they form `monadic' trees over $\Sigma\cup\{\bot\}$,
where $\rank(\sigma)=1$ for each $\sigma \in \Sigma$,
and $\rank(\bot)=0$. 

%% ===============================================
\para{Tree-walking automata}
For $k \ge 1$, a
 $k$-head  \emph{tree-walking automaton} 
is a finite-state automaton
equipped with $k$ heads
that walks on an input tree 
(over a given ranked alphabet $\Sigma$)
by moving its heads along the edges
from node to node%
\footnote{Maybe its heads should be called feet.}. 
At each moment it determines its next step
based on its present state, and the label and child number of
the nodes visited. 
Accordingly, it changes state and, 
for each of its heads, it stays at the node,
or moves either up to
the parent of the node, or down to a specified child.
If the automaton has no next step, we say it
\emph{halts}.

The \emph{language $L(\cA) \subseteq T_\Sigma$ accepted} by the 
$k$-head 
tree-walking automaton $\cA$ is the set of all trees 
over $\Sigma$
on which $\cA$ has a
computation starting with all its heads
at the root of the tree in the
initial state and halting in an accepting state,
again with all heads at the root of the tree.
The family of languages accepted by $k$-head deterministic
tree-walking automata is denoted by 
$\DTWA{k}$, for nondeterministic automata we write 
$\NTWA{k}$.

Such an automaton is able to make a systematic
search of the tree (which can be tuned to be, e.g.,  
a preorder traversal), even using a single head, as follows.
When a node is reached for the first time
(entering it from above) the automaton continues 
in the direction of the first child;
when a leaf is reached, the automaton goes up again.
If a node is reached from below, from a child,
it goes down again, to the next child, if that exists;
otherwise the automaton continues to the parent of the
node.
The search ends when the root is entered from its last
child.
This traversal is often used in constructions in this paper.

In both \cite{NevSch} and \cite{OkhSalDom}, as an example,
the authors explicitly construct a deterministic 
$1$-head
tree-walking automaton that evaluates boolean trees,
i.e., terms with binary operators `$\mbox{and}$' and
`$\mbox{or}$' and constants $0$ and $1$. 

Again, strings form a special case.
Tree-walking automata on monadic trees are 
equivalent to the usual two-way automata on strings.
A tree-walking automaton is able to recognize the
root of a tree as well as its leaves
(using child number and rank of the symbols).
This corresponds to a two-way automaton moving on a tape,
where the input string is written with two endmarkers
so the automaton knows the beginning and end of its input.

%% ===============================================
\para{Logic for Trees}
For an overview of the theory of first-order and monadic second-order
logic on both finite and infinite strings and trees
in relation to formal language theory, see \cite{Tho}.

In this paper our primary interest is in
\emph{first-order} 
logic, describing properties of trees. 
The logic has node variables $x,y,\dots$, which 
for a given tree range over its nodes. 
There are four types of atomic formulas over $\Sigma$: 
  $\lab_\sigma(x)$, for every $\sigma \in \Sigma$, 
meaning that $x$ has label $\sigma$; 
  $\edg_i(x,y)$, for every 
$i$ at most the rank of a symbol in $\Sigma$, 
meaning that the $i$-th child of $x$ is $y$; 
  $x \leq y$, meaning that $x$ is an ancestor of $y$; 
and 
  $x = y$, with obvious meaning.
The formulas are built from the atomic formulas using the
connectives  $\neg$, $\wedge$, and $\vee$, 
as usual; variables 
can be quantified with  $\exists$ and $\forall$.

If $t$ is a tree over $\Sigma$,
$\phi$ is a formula over $\Sigma$ 
such that its free variables are $x_1,\dots, x_n$,
and $u_1,\dots, u_n$ are nodes of $t$,
then we write $t \models \phi(u_1,\dots, u_n)$
if formula $\phi$ holds for $t$ where the free $x_i$ are valuated
as $u_i$.

For fixed $k\ge 1$, by overlined symbols like $\vec{x}$
we denote $k$-tuples of objects of the type referred to by $x$,
like logical variables, nodes in a tree, or pebbles used by 
an automaton.
By $x[i]$ we then denote the $i$-th component of $\vec{x}$.

We consider the additional operator of $k$-ary
\emph{transitive closure}.
Let $\phi(\vec{x},\vec{y})$ be a formula where 
$\vec{x},\vec{y}$ are $k$-tuples of distinct variables occurring
free in $\phi$.
We use $\phi^*(\vec{x},\vec{y})$ to denote the 
transitive closure of $\phi$ with respect to $\vec{x},\vec{y}$.
Informally, 
$\phi^*(\vec{x},\vec{y})$ 
means that we can make a series of jumps
from $k$-tuple $\vec{x}$ to $k$-tuple $\vec{y}$ such that
each pair of consecutive $k$-tuples $\vec{x}',\vec{y}'$
connected by a jump satisfies 
$\phi(\vec{x}',\vec{y}')$.
More formally, 
let $\phi$ have $2k+m$ free variables 
$\vec{x},\vec{y},z_1,\dots,z_m$.
For tree $t$ and nodes $\vec{u},\vec{v},w_1,\dots,w_m$ of $t$ we have
$t \models \phi^*(\vec{u},\vec{v},w_1,\dots,w_m)$ 
if there exists a sequence of $k$-tuples of nodes
$\vec{u}_0,\vec{u}_1, \dots, \vec{u}_n$, $n\ge 0$,
such that
$\vec{u}=\vec{u}_0$, $\vec{v}=\vec{u}_n$, and 
$t \models \phi(\vec{u}_i,\vec{u}_{i+1},w_1,\dots,w_m)$ 
for each $0\le i < n$.
In particular, $t \models \phi^*(\vec{u},\vec{u},w_1,\dots,w_m)$
for every $k$-tuple $\vec{u}$ of nodes of $t$.

Formally we should specify the $k$-tuples
 $\vec{x},\vec{y}$,
or rather $\vec{x}',\vec{y}'$,
of free variables
with respect to which to take the transitive closure, like for
the usual universal and existential quantification, and
write $(\tc(\vec{x}',\vec{y}')
\phi(\vec{x}',\vec{y}'))(\vec{x},\vec{y})$ instead of 
$\phi^*(\vec{x},\vec{y})$.

A predicate $\phi(\vec{x},\vec{y})$ with free variables 
$\vec{x},\vec{y}$ is
\emph{functional} (in $\vec{x},\vec{y}$) 
if for every tree $t$ and $k$-tuple of nodes $\vec{u}$ 
there is at most one $k$-tuple 
$\vec{v}$ such that $t \models \phi(\vec{u},\vec{v})$.
If $\phi$ has more free variables than $\vec{x},\vec{y}$,
this should hold for each fixed valuation of those variables.
The transitive closure $\phi^*(\vec{x},\vec{y})$ 
is \emph{deterministic} if $\phi$ is functional
(in the variables with respect to which the transitive closure
is taken).
Instead of requiring $\phi$ to be functional we could, equivalently,
require it to be of the form 
$\psi(\vec{x},\vec{y}) \land \forall \vec{z}(\psi(\vec{x},\vec{z})\to
\vec{y}=\vec{z})$.
This has the advantage of being a decidable property, but it is less 
convenient in proofs.
%
%Note that functionality of a predicate
%$\phi(x,y)$ is decidable, as one easily writes
%a formula in monadic second-order logic that expresses this property.

\smallskip
The \emph{tree language} defined by a closed 
formula $\phi$ over $\Sigma$ 
consists of all trees  $t$
in $T_\Sigma$ such that $t \models \phi$.
The family of all tree languages that are first-order definable
is denoted by $\FO$;
if one additionally allows $k$-ary transitive closure or 
deterministic transitive closure 
we have the families $\FOTC{k}$ and $\FODTC{k}$, respectively.
General transitive closure and deterministic transitive closure
(i.e., over unbounded values of $k$) 
characterize the complexity classes 
\NLOG\ and \DLOG\ 
 (for strings, or more generally, ordered structures),
respectively,
see \cite{EbbFlu,Imm99}. 

By $\LFO$ we denote the family of tree languages definable
in \emph{local} first-order logic, i.e., 
dropping the atomic formula $x\le y$. 
One should note however, that $x\le y$ is the
transitive closure of the functional predicate 
`$x$ parent of $y$', i.e.,  $\bigvee_i \edg_i(x,y)$.
Hence $x\le y$ is expressible in $\mbox{LFO+DTC$^1$}$,
and the families $\FODTC{k}$, etc., 
of tree languages definable in first-order logic with
transitive closure,
do not change by this restriction.

In Section~\ref{overview} we study the specific case $k=1$, i.e.,
we consider unary transitive closure only. 
The family \FOTC1 is included in the family of regular tree
languages, i.e., the 
tree languages that are definable in
\emph{monadic second-order} logic,
which additionally has 
node set variables $X,Y,\dots$,
ranging over sets of nodes of the tree;
it allows quantification over these variables, and 
has the predicate $x \in X$, 
with its obvious meaning.

\medskip
%% ===============================================
\begin{example}\label{walking}
It is proved in \cite{BoCo2} 
that there is a regular tree language $T$ that cannot be accepted
by any single-head nondeterministic tree-walking automaton.
Here we illustrate how to construct an 
\logic{FO+DTC$^1$} formula for that language.

Let $\Sigma =\{a,b,c\}$,
where $a$ and $b$ are nullary 
(labelling, of course, the leaves of the trees over $\Sigma$) 
and $c$ is binary (labelling the internal nodes). 
The language $T$
consists of all trees over $\Sigma$ for which the path to each
leaf labelled by $a$ contains an even number of 
`branching' nodes, i.e., internal nodes for which both
the left and right subtree contain an $a$-labelled leaf.

It is easy to construct a first-order formula expressing 
that a node is branching. Let $\psi(x,y)$ specify
that $y$ is the lowest branching ancestor of $x$, 
and let $\phi(x,y) \equiv (\exists z)( \psi(x,z)\land \psi(z,y))$
to claim $y$ is the second lowest branching ancestor of $x$.
Observe that $\phi$ is functional.
Now $T$ is specified by the \logic{FO+DTC$^1$} formula 
$(\forall x)( \lab_a(x) \to 
(\exists y)[\; \phi^*(x,y) \land \lnot(\exists z)\psi(y,z)  \;]\; )$.

In fact $T$ even belongs to $\FO$, as observed in \cite{BoCo2},
but earlier in \cite[Lemma~5.1.8]{Pott}.
\qed
\end{example}
%% ===============================================

%% ===============================================
%% ===============================================
\section{Tree-Walking Automata with Nested Pebbles}
%% ===============================================

A $k$-head tree-walking automaton with \emph{nested pebbles} is a
$k$-head tree-walking automaton that is 
additionally equipped with a finite set of
pebbles. During the computation it may drop these pebbles 
(one by one)
on
nodes visited by its heads,
to mark specific positions.
It may test the currently visited
 nodes to see which pebbles are present.
Moreover, it may retrieve a pebble from anywhere in the
tree, provided the \emph{life times} of the pebbles are nested.
This can be formalized by keeping a (bounded) 
stack in the configuration of the automaton, pushing and
popping pebbles when they are dropped and retrieved,
respectively.
Note that since this stack is bounded by the number of pebbles,
it can also be kept in the finite control of the automaton.%
\footnote{%
If the automaton uses pebbles $x_1,\dots, x_n$, then the
contents of the stack can be any string over $\{x_1,\dots, x_n\}$
in which each $x_i$ occurs at most once.
It can be assumed w.l.o.g. that the stack always contains
$x_1 x_2\cdots x_i$ for some $i$, but we will do this only in the proof of
Lemma~\ref{DPTWA->FODTC}
(where, in fact, the order is reversed).}
Pebbles can be reused any number of times (but there is
only one copy of each pebble).
Accepting
computations should start and end with all heads at the root
without pebbles on the input tree.

The family of tree languages accepted by deterministic $k$-head
tree-walking automata with nested pebbles is denoted by
$\DPTWA{k}$, the nondeterministic variant by 
$\NPTWA{k}$.

\smallskip
Some specific properties of these pebbles must be stressed.
First, 
as stated above, pebbles are used in a {\sc lifo} manner,
as in \cite{GloHar,walking,MilSV,Via,EngMan,NevSchVia},
which means that only the last one dropped can be retrieved, and
thus
their life times on the tree are nested.
Without this restriction again the classes \DLOG\ and \NLOG\
would be obtained.
Second, 
this is rather nonstandard,
the automaton need not return one of its heads 
to the position where
a pebble was dropped in order to pick it up: at any moment
the last pebble dropped can be retrieved.
This means that the pebble behaves as a \emph{pointer}:
we can store the address of a node when we know it
(which is the case when we visit it)
and we can later wipe the address from memory without the need to
return to the node itself.
Such pebbles were called ``abstract markers'' in \cite{BluHew}
 (to distinguish them from the usual ``physical markers'').
Finally, as opposed to \cite{GloHar},
during the computation all pebbles dropped remain visible to
the automaton (and not only the one or two on top of the stack).

%% ===============================================
\begin{example}\label{DPTWA}
The regular tree language $T$ from Example~\ref{walking}
cannot be accepted
by any single-head nondeterministic tree-walking automaton
(without pebbles), as proved in \cite{BoCo2}.
As an example, here
we show how to accept $T$
by a (single-head) deterministic
tree-walking automaton with two nested pebbles.

Using a preorder traversal of the input tree,
the first pebble is placed consecutively on leaves labelled by
$a$. For each such position, starting at the leaf we follow the path
upwards to the root counting the number of branching nodes.
To test whether an internal node is branching we place 
the second
pebble on the node and test whether its other subtree,
i.e., the subtree that does not contain the first pebble,
contains an $a$-labelled leaf
(using again a preorder traversal of that subtree,
the root of which can be recognized through the second pebble
which marks the parent of that root).
After testing the node, we pick up the second pebble.
At the root,
we reject whenever we count an odd number of such nodes on a
path;
otherwise we return to the position of the first pebble
(using another preorder traversal of the tree).

In fact, the language can be accepted with just one pebble%
\footnote{As brought to our attention by 
Christof L\"oding (and his student Gregor Hink).}
(which is nested trivially).
As explained above, the pebble can be used to detect
all branching nodes, which,
together with all $a$-labelled leaves, can be
viewed as a binary tree.
To check, for that tree, that the path to each leaf is of even
length, the automaton performs a preorder traversal and counts
its number of steps, modulo $2$.
At each leaf the count should be $0$.
\qed
\end{example}

\noindent
To fix the model, a $k$-head tree-walking pebble automaton 
is specified as a tuple
$\cA = (Q,\Sigma, X, q_0, A, I)$,
where
$Q$ is a finite set of states,
$\Sigma$ is a (ranked) input alphabet,
$X$ a finite set of pebbles,
$q_0\in Q$ the initial state,
$A\subseteq Q$ the set of accepting states,
and $I$ the finite set of instructions.

Each instruction is a triple of the form 
$\tuple{p,\psi,q}$
or
$\tuple{p,\varphi,q}$
or
$\tuple{p,\negg\varphi,q}$,
where $p,q\in Q$ are states,
$\psi$ is an operation, and $\varphi$ a test.
There are four types of operations: 
$\op{up}_i$, $\op{down}_{i,j}$
(moves),
$\op{drop}_i(x)$, and $\op{retrieve}(x)$
(pebble operations),
and three types of tests: 
$\op{lab}_{i,\sigma}$,
$\op{peb}_i(x)$, and
$\op{chno}_{i,j}$,
where in each case
$i$ indicates a head
($1\le i\le k$),
$j$ is a child number
($1\le j\le \max\{\mbox{rank}(\sigma)\mid \sigma\in\Sigma\} $),
$\sigma\in\Sigma$ is a node label,
and
$x\in X$ is a pebble.
An instruction $\tuple{p,\chi,q}$ is called an outgoing
instruction of state $p$.

The automaton $\cA$ is deterministic if for any pair
$\tuple{p,\chi_1,q_1}$,
$\tuple{p,\chi_2,q_2}$
of distinct instructions starting in the same state,
either $\chi_1=\negg\chi_2$
or $\chi_2=\negg\chi_1$.

A configuration of $\cA$ on tree $t$ over $\Sigma$
is a triple $[p, \vec{u}, \alpha]$,
where $p\in Q$ is a state,
$\vec{u}$ is a $k$-tuple of nodes of $t$ indicating
the positions of the $k$ heads,
and $\alpha = (x_1,w_1)\cdots(x_m,w_m)$
the stack of pebbles dropped at their positions
($m\ge 0$, $x_j\in X$, $w_j$ a node of $t$).
The initial configuration equals
$[ q_0, \vec{\op{root}},\varepsilon ]$,
where $\vec{\op{root}}$ consists of $k$ copies of the
root of $t$,
and $\varepsilon$ is the empty stack.

The semantics of the pebble automaton is defined using
the step relation $\vdash_{\cA,t}$ on configurations
for automaton $\cA$ on input tree $t$.
We have
$[p,\vec{u},\alpha] \vdash_{\cA,t} [q,\vec{v},\beta]$ 
with
$\alpha = (x_1,w_1)\cdots(x_m,w_m)$,
if there exists an instruction
$\tuple{p,\chi,q}$ such that
\[
\begin{array}{rll}
& \mbox{if}  & \mbox{then}
\\
\chi = {}& \op{up}_i 
&
v[i] \mbox{ is the parent of } u[i],
v[h] = u[h] \mbox{ for } h\neq i, \alpha = \beta
\\
&
\op{down}_{i,j}
&
v[i] \mbox{ is the $j$-th child of } u[i],
v[h] = u[h] \mbox{ for } h\neq i, \alpha = \beta
\\
&
\op{drop}_i(x)
&
\beta = \alpha(x,u[i]), 
x\notin\{x_1,\dots,x_m\}, \vec{u}=\vec{v}
\\
&
\op{retrieve}(x)
&
m\ge 1, 
\beta(x_m,w_m) = \alpha, x=x_m,
\vec{u}=\vec{v}
\\
&
\op{lab}_{i,\sigma}
&
u[i] \mbox{ has label } \sigma, \vec{u}=\vec{v}, \alpha = \beta
\\
&
\op{peb}_i(x)
&
(x,u[i]) \mbox{ occurs in } \alpha, \vec{u}=\vec{v}, \alpha = \beta
\\
&
\op{chno}_{i,j}
&
\mbox{the child number of } u[i] \mbox{ is } j,
\vec{u}=\vec{v}, \alpha = \beta
\end{array}
\]
or in case of the negative tests,
$\chi = \negg\op{lab}_{i,\sigma},
\negg\op{peb}_i(x), \negg\op{chno}_{i,j}$,
the tests above are negated, whereas head positions and
pebble stack remain unchanged ($\vec{u}=\vec{v}, \alpha = \beta$).

A configuration $c$ is halting if there is no $c'$ such that
$c \vdash_{\cA,t} c'$, and it is accepting if it is halting
and $c= [p,\vec{\op{root}},\varepsilon]$ for some $p\in A$.
Then the language accepted by $\cA$ is defined as
$L(\cA)=
\{\; t\in T_\Sigma \mid
[q_0,\vec{\op{root}},\varepsilon] \vdash^*_{\cA,t} 
c$ for some accepting configuration $c \;\}$.

%% ===============================================
\begin{example}
Let $\Sigma$ be as in Example~\ref{walking}.
We write, in our formalism, a deterministic
single-head 
tree-walking automaton $\cA$ (without pebbles)
such that $L(\cA)$ consists of all trees over $\Sigma$
that have $a$-labelled leaves only.
The automaton performs a preorder traversal of the input tree.
The main states are as follows.
In state $1$ we move down to the left child 
until we reach a leaf,
in state $2$ we are at a left child and move to its right sibling,
and in state $3$ we move up until we are at a left child
or at the root.

As the automaton has only a single head,
we omit the head number in the instructions.
The initial state is $1$, the only accepting state is $h$.
The automaton has the following instructions:

$(1,\op{lab}_{c},1')$,
$(1',\op{down}_1,1)$,
$(1,\negg\op{lab}_{c},1'')$,
$(1'',\op{lab}_{a},3)$,

$(3,\op{chno}_2,3')$,
$(3',\op{up},3)$,
$(3,\negg\op{chno}_2,3'')$,

$(3'',\op{chno}_1,2)$,
$(3'',\negg\op{chno}_1,h)$,

$(2,\op{up},2')$,
$(2',\op{down}_2,1)$.
\qed
\end{example}
%% ===============================================

%% ===============================================
%% ===============================================
\section{From Logic to Nested Pebbles}
\label{logpeb}
%% ===============================================

\medskip 
We now generalize the inclusion $\FO \subseteq \DPTWA{1}$ 
from Section~5 of \cite{walking},
on the one side introducing $k$-ary transitive closure,
on the other side allowing $k$ heads. 
Note also the result is for the `pointer' variant of pebbles,
rather than pebbles that have to be picked up where they were
dropped.
%$\DPTWA \subseteq \REGT$ 

%% ===============================================
\begin{lemma}\label{FODTC->DPTWA}
For trees over a ranked alphabet, and $k \ge 1$, 
$\FODTC{k} \subseteq \DPTWA{k}$.
\end{lemma}

\proof
The proof is by induction on the structure of the formula.
For each first-order formula with deterministic transitive
closure we construct a deterministic tree-walking automaton
with nested pebbles that always halts
(with all its heads at the root).
The additional effort we have to take to make it always halting,
will pay itself back when we deal with negation, but is also
helpful when considering disjunction and 
existential quantification.
Generally speaking,
each variable of the formula acts as a pebble for the automaton.
In case of $k$-ary transitive closure we need $3k$ pebbles
to test the formula by an automaton.
Most features can be simulated using a single head,
moving pebbles around the tree, 
only for transitive closure we need all the $k$ heads.

As intermediate formulas may have free variables we
need to extend our notion of recognizing a tree by an
automaton: a valuation of the free variables is fixed
by putting pebbles on the tree, one for each variable,
and the automaton should
evaluate the formula according to this valuation.

More formally, 
let $\phi = \phi(x_1,\dots,x_n)$ be a formula with free
variables  $x_1,\dots,x_n$. The automaton $\cA$ for $\phi$ 
should check whether 
$t \models \phi(u_1,\dots,u_n)$ for 
nodes $u_1,\dots,u_n$ in a tree $t$, as follows. 
It is started in the initial state 
with all heads at the root of the tree
$t$, where $u_1,\dots,u_n$  are marked with pebbles
$x_1,\dots,x_n$. During the computation $\cA$ may use
additional pebbles (in a nested fashion) and it may test 
$x_1,\dots,x_n$, but it is not allowed to retrieve them.
The computation should halt 
again with all heads 
at the root of $t$ with the
original configuration of pebbles. 
The halting state is accepting if and only if 
$t \models \phi(u_1,\dots,u_n)$.

\medskip
For the atomic formulas it is straightforward to construct
(single-head) automata.
As an example, for 
$\phi = x\le y$ the automaton searches the tree
for a marked node representing $y$. From that position 
the automaton walks upwards to the root, where it halts,
signalling whether $x$ was
found on the path from $y$ to the root.
For $\edg_i(x,y)$ the automaton searches for $x$, then
determines whether $x$ has an $i$-th child
(the arity of the node can be seen from its label)
and moves to that child.
There the automaton checks whether pebble $y$ is present.

For the negation $\phi = \lnot\phi_1$ 
of a formula we use the original automaton for $\phi_1$, 
but change its accepting states to the complementary set. 
This construction works thanks to
the fact that the automata we build are always halting.

A similar argument works for the conjunction
$\phi = \phi_1 \land \phi_2 $, 
or the disjunction
$\phi = \phi_1 \lor \phi_2 $,
of two formulas. We may run the two automata constructed
for the two constituents consecutively. Note that
the free variables in $\phi_1$ and $\phi_2$ need
not be the same, but the extra pebbles that need to
be present for $\phi$ are ignored by each of
the two automata.

\smallskip 
For quantification 
$\phi = (\forall x) \phi_1$
the automaton makes a systematic traversal through the tree,
using a single head.
When it reaches a node (for the first time) it places
a pebble $x$ at that position.
Then it returns to the root, and runs the automaton
for $\phi_1$ as a subroutine;
the free variable $x$ of $\phi_1$ is marked by the pebble,
as requested by the inductive hypothesis.
When this test for $\phi_1(x)$ is positive,
i.e., the subroutine 
halts at the root in an accepting state,
the automaton returns to the node marked 
$x$  (by searching for it in the tree),
picks up the pebble, and places it on the next node
of the traversal.
When the automaton has successfully run the test 
for $\phi_1$ for each node it accepts.
The formula $(\exists x) \phi_1$ is treated similarly.

\smallskip 
The main new element of this proof compared to 
\cite{walking}
is the introduction of $k$-ary transitive closure.
Here we need to program a walk from one 
$k$-tuple of nodes to another $k$-tuple
with `jumps' specified by a $2k$-ary formula.
We cannot do this in a straightforward way, as we might
end `jumping around' in a cycle without noticing.
Such an infinite computation violates the requirement that
our automaton should always halt. 
We use a variant of the technique of
Sipser \cite{Sip} to avoid this trap,
and run this walk backwards.

So, let $\phi = \phi^*_1$ be the transitive
closure of a  functional $2k$-ary predicate
$\phi_1$. Given a tree with $2k$ nodes marked by pebbles
$\vec{x}$ and $\vec{y}$ we have to construct an automaton 
$\cA$ that
decides whether we can connect the $k$-tuples $\vec{x}$
and $\vec{y}$ by a series of
intermediate $k$-tuples such that $\phi_1$ holds for each
consecutive pair.
In what follows we assume that $\phi_1$ has $2k$ free variables
$\vec{x}$ and $\vec{y}$ 
(with respect to which the transitive closure is taken),
and we disregard the remaining free variables of $\phi_1$
(the values of which are fixed by pebbles).

\newcommand{\Tree}{$t_k(\vec{v})$}
% yes! k-ary (oorspronkelijk met phi-1 corresponderend)
Now consider the set of $k$-tuples of 
nodes of the input tree $t$, 
spanning the (virtual) computation space of $\cA$.
We build a directed 
graph on these $k$-tuples by connecting
vertex%
\footnote{%
For clarity we distinguish `node' in the input tree
from `vertex' in the computation space, i.e., a $k$-tuple of nodes.
Similarly we use `edge' and `arc'.
}
 $\vec{u}$ to vertex $\vec{v}$ if 
$t \models \phi_1(\vec{u},\vec{v})$, i.e., the pair 
$(\vec{u},\vec{v})$ in $t$ satisfies $\phi_1(\vec{x},\vec{y})$.
As $\phi_1$ is functional, for each vertex there is at most
one outgoing arc. 
Thus, if we fix a vertex $\vec{v}$ (of $k$ nodes in $t$) 
and throw
away the outgoing arc of $\vec{v}$ (if it exists) the component
of vertices connected to $\vec{v}$ in this graph forms a tree
\Tree, 
with arcs pointing towards the root $\vec{v}$ rather
than towards the leaves. 
Note this is a directed tree in the 
graph-theoretical sense; there is no bound on the number of
arcs incident to each vertex.

Of course, this tree \Tree\ 
with $\phi_1$-arcs 
consists of all vertices $\vec{u}$ 
that satisfy $t \models \phi(\vec{u},\vec{v})$, 
and fixing $\vec{v}$ to be
the vertex marked by the pebbles $\vec{y}$ 
the new automaton $\cA$ traverses that tree  \Tree\
and tries
to find the vertex marked by pebbles $\vec{x}$. 

However, the tree \Tree\ is not explicitly available,
and has to be reconstructed while walking on the input
tree $t$, using the automaton $\cA_1$ for $\phi_1$ as a
subroutine.
In particular, we want to implement a traversal on \Tree.
As the vertices of \Tree\ consist of $k$-tuples of nodes
of $t$, we order these $k$-tuples in a natural way using the 
lexicographical ordering based on the 
preorder in $t$. In this way we impose an ordering on
the children of each vertex of \Tree,
thus allowing the usual preorder traversal of \Tree\ as described below.
To find the successor of a $k$-tuple $\vec{z}$ in the
lexicographical ordering we act like adding one to a 
$k$-ary number:
change the last coordinate of the tuple $\vec{z}$ into its
successor 
(here the preorder successor in $t$)
if that exists, otherwise reset that coordinate
to the first element (here the root of $t$),
and consider the last-but-one coordinate, etc.

We implement a preorder traversal of  \Tree, which means
we compute
the preorder successor of each vertex of the tree
\Tree\ whose arcs are defined by $\phi_1$, and 
where the ordering between sibling vertices is based on the
lexicographical ordering of nodes in $t$:

%\newpage
\noindent
\begin{tabbing}
xxxx\=xxxx\=xxxx\=\kill
\>\+preorder successor of vertex $\vec{u}$ in \Tree:\\
\>\+if it exists, the first child of $\vec{u}$,\\
else, on the path of $\vec{u}$ to the root $\vec{v}$,\\
\>the right sibling of the first vertex that has one.
\end{tabbing}

We traverse the tree \Tree, with  $2k$ pebbles $\vec{x}$ and
$\vec{y}$ fixed, with the help of $3k$ additional pebbles 
$\vec{x}'$, $\vec{y}'$, and $\vec{z}'$. 
During this traversal, $\cA$ keeps track of the current vertex
of \Tree\ with its $k$ heads.
Initially the heads move to $\vec{y}$,
i.e., to $\vec{v}$.
Note that the order of dropping the pebbles $\vec{x}'$ and $\vec{y}'$
differs in the two cases below: in the first case we have to 
check $\phi_1(\vec{x}',\vec{y}')$ `backwards', finding 
$\vec{x}'$ given $\vec{y}'$, while in the second case it is the
other way around. This is reflected in the order of dropping 
$\vec{x}'$ and $\vec{y}'$.

First, we describe how to check whether the current vertex
has a first child in \Tree, and to go there if it exists.
We drop pebbles $\vec{y}'$ to fix the current vertex, and 
we systematically place pebbles $\vec{x}'$ on
each candidate vertex, i.e., each
$k$-tuple of nodes of the tree $t$ (except $\vec{v}$). 
Thus, lexicographically, in each step 
the last pebble of $\vec{x}'$ is carried to the next node in 
$t$ (with respect to the preorder in $t$), 
but when that pebble has been at all nodes, it is lifted,
the last-but-one pebble is moved to its successor node in $t$, 
and the
last pebble is replaced on the root, etc.
For each $k$-tuple $\vec{x}'$ we check
$\phi_1(\vec{x}',\vec{y}')$ using automaton $\cA_1$ as a subroutine.
If the formula is true, we have found the first child in \Tree\,
and we move the $k$ heads
to the nodes marked by $\vec{x}'$, 
lift pebbles $\vec{x}'$,
and retrieve pebbles $\vec{y}'$ (from a distance). 
If the formula is not true, we move
$\vec{x}'$ to the next candidate vertex as described above
(but $\vec{v}$ is disregarded).
If none of the candidates $\vec{x}'$ satisfies 
$\phi_1(\vec{x}',\vec{y}')$, the
vertex $\vec{y}'$ obviously has no child in  \Tree.

Second, we describe how to check for a right sibling
in \Tree, and
go there if it exists, or go up (to the parent of the
current vertex) otherwise. The problem here is to keep the
pebbles in the right order, adhering to the nesting of the
pebbles.
First drop pebbles $\vec{x}'$ on the current vertex. Then
determine its parent in \Tree; this is the
unique vertex that satisfies $\phi_1(\vec{x}',\vec{y}')$, 
where $\vec{y}'$
marks the parent vertex of $\vec{x}'$, thanks to the
functionality of $\phi_1$. It can be found in a
traversal of all $k$-tuples of nodes of $t$ 
using pebbles $\vec{y}'$ and subroutine $\cA_1$
(as described above for the first child).
Leave $\vec{y}'$ on the parent and return to $\vec{x}'$ 
(by searching
for $\vec{x}'$ in the tree $t$). 
Using the third set of $k$ pebbles $\vec{z}'$,
traverse the $k$-tuples of nodes of 
$t$ from $\vec{x}'$ onwards and try to find the next $k$-tuple
that satisfies $\phi_1(\vec{z}',\vec{y}')$ 
when $\vec{z}'$ is dropped. If
it is found, it is the right sibling of $\vec{x}'$.
Return there, lift $\vec{z}'$,
and retrieve $\vec{y}'$ and $\vec{x}'$. 
If no such $k$-tuple is found, the current
vertex has no right sibling, and we go up in the tree 
\Tree, i.e., we return to $\vec{y}'$. 
Here we lift $\vec{y}'$ and retrieve $\vec{x}'$.

In all these considerations special care has to be taken
of the root $\vec{v}$. It has no parent 
in \Tree. 
Fortunately $\vec{v}$ is clearly marked by pebbles $\vec{y}$.
\qed

\noindent
The number of pebbles needed to compute a formula of
\logic{FO+DTC$^{k}$}
 according to the construction above depends
only on the nesting of quantifiers and transitive closures
in the formula. For each quantifier we count a single pebble,
and $3k$ for transitive closure, and compute the maximum needed
over all sequences of nested operators in the formula.

\medskip
When
allowing transitive closure of arbitrary formulas
(not requiring them to be functional)
it is customary to restrict attention to
formulas with only \emph{positive occurrences} of 
transitive closure, i.e., within the scope of an even number of
negations (see, e.g., \cite{EbbFlu,Imm99}). 
Using standard argumentation each such formula is equivalent 
to one where 
negation is applied to atomic formulas only.

For such formulas there is a similar, nondeterministic, result
as the one above.
Atomic formulas and their negations are treated as above, 
and so are conjunction and universal quantification.
For disjunction and existential quantification, 
the automaton uses nondeterminism in the obvious way. 
For transitive closure, the Sipser technique we have used 
in the previous proof is not needed.
For a formula $\phi = \phi^*_1$ 
the automaton $\cA$ checks nondeterministically 
the existence of a path $\vec{u}_0,\vec{u}_1,\dots,\vec{u}_n$ 
from vertex $\vec{x}$ to vertex $\vec{y}$
in the directed graph determined by $\phi_1$
(described in the proof of Lemma~\ref{FODTC->DPTWA}). 
When $\cA$ is at vertex $\vec{u}_i$, it proceeds to vertex $\vec{u}_{i+1}$
using $2k$ additional pebbles $\vec{x}'$ and $\vec{y}'$, as follows.
It drops $\vec{x}'$ on the current nodes $\vec{u}_i$ 
and nondeterministically chooses nodes $\vec{u}_{i+1}$,
where it drops $\vec{y}'$ and checks that $\phi_1(\vec{x}',\vec{y}')$.
Then it returns to $\vec{y}'$, lifts $\vec{y}'$, and retrieves $\vec{x}'$. 

We denote the positive restriction of 
$\FOTC{k}$ by $\FOpos{k}$, and similarly for the deterministic case. 
Thus, 
for trees over a ranked alphabet, nondeterministic
$k$-head 
tree-walking automata can compute
positive $k$-ary transitive closure:
$\FOpos{k} \subseteq \NPTWA{k}$.

%% ===============================================
%% ===============================================
\section{From Nested Pebbles to Logic}
\label{peblog}
%% ===============================================

The classical result of Kleene \cite{Kle} shows how to transform a
finite-state automaton into a regular expression,
which basically means that we have a way to
dispose of the states of the automaton.
Bargury and Makowsky \cite{BarMak} 
observe that this technique can also be used to
transform multi-head automata walking on grids into
equivalent formulas with transitive closure:
transitive closure may very well specify 
sequences of consecutive positions
on the input, but has no direct means to store states.  
A similar technique is used here.
As our model includes pebbles, this imposes an additional
problem, which we solve by iterating the construction 
for each pebble.
Unlike \cite{BarMak} we have managed to find a formulation that
works well for both the nondeterministic and deterministic case.

\smallskip
Given a (deterministic) computational finite-state
device with $k$ heads on the tree, the 
step relation of which is 
specified by logical formulas, 
we show that the computation relation
that iterates consecutive steps
can be expressed using $k$-ary (deterministic) transitive closure.
Of course, the consecutive positions of the heads along the tree 
are well taken care of by the closure operator, but here we
additionally require that the states of the device 
should match the sequence of steps.

Let $\Phi$ be a $Q\times Q$ matrix of predicates
$\phi_{p,q}(\vec{x},\vec{y})$, $p,q\in Q$
for some finite set $Q$ (of states),
where $\vec{x},\vec{y}$ each are $k$ distinct variables occurring 
free in all 
$\phi_{p,q}$.
We define the \emph{computation closure} 
of $\Phi$ with respect to $\vec{x},\vec{y}$ as the matrix
$\Phi^\# $ consisting of predicates 
$\phi^\# _{p,q}(\vec{x},\vec{y})$
where 
$t \models \phi^\# _{p,q}(\vec{u},\vec{v})$ 
iff there exists a sequence
of $k$-tuples of nodes $\vec{u}_0, \vec{u}_1, \dots , \vec{u}_n$
and a sequence of states
$p_0, p_1, \dots , p_n$,
%$n \ge 0$,
$n \ge 1$,
such that 
$\vec{u} = \vec{u}_0$, $\vec{v} = \vec{u}_n$, 
$p = p_0$, $q = p_n$, 
where
%$p_i \neq q$ for  $1 \le i < n$,
%and 
$t \models \phi_{p_{i},p_{i+1}}(\vec{u}_{i},\vec{u}_{i+1})$ 
for $0 \le i < n$.
%and there exist no $r\in Q$ and no node $w$
%such that $t \models \phi_{q,r}(v,w)$.
(\footnote{%
Note that,
to simplify the description of computation closure we have 
disregarded the remaining free variables of the $\phi_{p,q}$
and $\phi^{\#}_{p,q}$.
More precisely, if $z_1,\dots, z_m$ are all the free variables of
all $\phi_{p,q}$ (in addition to $\vec{x},\vec{y}$),
then each $\phi^{\#}_{p,q}$ has free variables 
$\vec{x},\vec{y},z_1,\dots, z_m$.
In the definition of 
$t \models \phi^\# _{p,q}(\vec{u},\vec{v},w_1,\dots,w_m)$ 
the $z_1,\dots, z_m$ have fixed values $w_1,\dots, w_m$.%
})

Intuitively $t \models \phi^\# _{p,q}(\vec{u},\vec{v})$
means that there is a 
$\Phi$-path of consecutive steps
(as specified by matrix $\Phi$) leading from 
nodes $\vec{u}$ in state $p$ to nodes $\vec{v}$ in state $q$. 
Note that only nonempty paths are considered ($n\ge 1$). 

We say that $\Phi$ is \emph{deterministic} if
its predicates are both functional and
exclusive, i.e.,
for any $p,q,q' \in Q$ and $3k$ nodes
$\vec{u},\vec{v},\vec{v}'$ of any tree $t$,
if both $t \models \phi_{p,q}(\vec{u},\vec{v})$ 
and  $t \models \phi_{p,q'}(\vec{u},\vec{v}')$ 
then  $q=q'$ and $\vec{v}=\vec{v}'$.
Moreover, $\Phi$ is said to be \emph{semi-deterministic} 
if the previous requirement holds for final states $q,q'$ only, 
where $q$ is \emph{final} if $\phi_{q,r}$ is false for all $r\in Q$
(and similarly for $q'$). 

%% ===============================================
\begin{lemma}\label{Phi}\leavevmode
\begin{enumerate}
\item
If $\Phi$ is deterministic, then $\Phi^\#$ is semi-deterministic.
\item
If $\Phi$ is in $\mbox{\rm FO+TC$^k$}$, then so is $\Phi^\#$.
\item
If $\Phi$ is in \mbox{\rm FO+DTC$^k$} and deterministic,
then $\Phi^\#$ is in \mbox{\rm FO+DTC$^k$}.
\end{enumerate}
\end{lemma}

\proof
\emph{1.}
Let $q,q'\in Q$ be final, and 
assume that for tree $t$ both 
$\phi^\# _{p,q}(\vec{u},\vec{v})$ 
and 
$\phi^\# _{p,q'}(\vec{u},\vec{v}')$ 
hold for $p \in Q$ and $3k$ nodes
$\vec{u},\vec{v},\vec{v}'$ of $t$,
with $\Phi$-paths of length $n$ and $n'$ as
in the definition of computation closure ($n,n' \ge 1$).

Consider the first steps of both paths.
We have 
$\phi_{p,p_1}(\vec{u},\vec{u}_1)$ 
and 
$\phi_{p,p'_1}(\vec{u},\vec{u}'_1)$,
as well as
$\phi^\# _{p_1,q}(\vec{u}_1,\vec{v})$ if $n\ge 2$,
and 
$\phi^\# _{p'_1,q'}(\vec{u}'_1,\vec{v}')$ if $n'\ge 2$.
Due to the determinism of $\Phi$
we conclude $p_1 = p'_1$ and $\vec{u}_1 = \vec{u}'_1$.

If $n=1$, then $\vec{u}_1 = \vec{v}$ and $p_1 = q$. Since $p'_1 = q$ is final, 
$\phi^\# _{p'_1,q'}(\vec{u}'_1,\vec{v}')$ is false. 
Hence $n'=1$ and $\vec{v}' = \vec{u}_1 = \vec{v}$ and $q' = p_1 = q$ as required.
For $n,n' \ge 2$
we continue inductively with $p_1$ and $\vec{u}_1$.

\medskip
\emph{2.}
The proof is a logical interpretation 
of the method of 
McNaugton and Yamada \cite{McNYa}.
%Let $q_1,\dots, q_n$ be an ordering of the states in $Q$.
Without loss of generality we assume that 
$Q = \{1,2,\dots,m \}$.
We show by induction on $\ell$ how to construct 
a matrix $\Phi^{(\ell)}$ of formulas $\phi^{(\ell)}_{p,q}$ 
in $\mbox{FO+TC$^k$}$ which are defined as  
$\phi^{\#}_{p,q}$,
except that the intermediate states
$p_1,\dots, p_{n-1}$ are chosen from $\{1,\dots, \ell \}$.
In particular, for $\ell =0$ no intermediate states are allowed,
whereas for $\ell =m$ all states are allowed, so we have
%$\phi^{(m)}_{p,q} = \phi^{\#}_{p,q}$.
$\Phi^{(m)} = \Phi^{\#}$.

For $\ell =0$, the length of the path is one.
This means that 
$\Phi^{(0)} = \Phi$.

Given $\Phi^{(\ell )}$ we obtain $\Phi^{(\ell +1)}$ as follows.
Assume $\phi^{(\ell +1)}_{p,q}(\vec{x},\vec{y})$ holds. 
Either there exists a $\Phi$-path 
that does not visit state ${\ell +1}$
(i.e., $p_i \neq \ell+1$ for all $0< i < n$ to be precise),
or this state is visited one or more times during the path.
In the former case  $\phi^{(\ell )}_{p,q}(\vec{x},\vec{y})$ holds,
in the latter case we have a path from 
state $p$ to state ${\ell +1}$, 
perhaps looping several times from
${\ell +1}$ back to itself, and finally there is a path from
state ${\ell +1}$ to state $q$. Neither of these paths contains
${\ell +1}$ as intermediate state, so in this case 
$\phi^{(\ell +1)}_{p,q}(\vec{x},\vec{y})$ postulates the existence of
intermediate nodes
$\vec{x}'$ and $\vec{y}'$ such that
\[\phi^{(\ell )}_{p,\ell +1}(\vec{x},\vec{x}') \land 
(\phi^{(\ell )}_{\ell +1,\ell +1})^*(\vec{x}',\vec{y}') 
\land
\phi^{(\ell )}_{\ell +1,q}(\vec{y}',\vec{y}).\]

\medskip
\emph{3.}
In the previous part of the proof transitive closure was
applied to predicates 
$\phi^{(\ell )}_{\ell +1,\ell +1}$.
However, determinism of $\Phi$ entails functionality of
predicates of the form
$\phi^{(\ell )}_{r,\ell +1}$, 
by an argument analogous to the one in 1. above.
Note that state ${\ell +1}$ need not be final, but the 
%(nonempty!) 
paths to state ${\ell +1}$ 
cannot be extended because (by definition of $\Phi^{(\ell )}$) state ${\ell +1}$
cannot be visited intermediately.
Hence, each transitive closure is applied to a functional predicate,
i.e., it is a deterministic transitive closure.
\qed

%% ===============================================
\begin{lemma}\label{DPTWA->FODTC}
For trees over a ranked alphabet, and $k \ge 1$, 
$\DPTWA{k} \subseteq \FODTC{k}$. 
\end{lemma}

\proof
Consider a tree-walking pebble automaton with
$k$ heads.
We assume that 
(1) accepting states have no outgoing instructions
(i.e., if $\tuple{p,\chi,q}$ is an instruction, then $p$ is not
accepting),
(2) the initial state is not accepting, and
(3) if 
there is an instruction
$(p,\mbox{drop}_i(x),q)$,
then there is no instruction $(q,\mbox{retrieve}(x),r)$.
The latter two requirements are to ensure 
that accepting computations,
and computations between dropping and retrieving a pebble,
are nonempty, allowing the use of Lemma~\ref{Phi}.

Let the automaton use $n$ pebbles,
$x_n, \dots, x_1$, where pebbles are placed on the tree in
the order given, i.e., $x_n$ is always placed on the bottom
of the pebble stack. 
We view the automaton as consisting of $n+1$
`levels' $\cA_n,\dots,\cA_1,\cA_0$
such that $\cA_\ell $ is a $k$-head 
tree-walking pebble automaton with $\ell $ pebbles 
$x_\ell , \dots, x_1$, available for dropping and retrieving,
whereas pebbles $x_n, \dots, x_{\ell +1}$ have a fixed
position on the tree and the automaton $\cA_\ell $ 
may test for their presence.
Basically,  $\cA_\ell $  acts as a
tree-walking automaton that drops pebble $x_\ell $,
then queries pebble automaton $\cA_{\ell -1}$ 
with $\ell -1$ pebbles
where to go in the tree, moves there, 
and retrieves pebble $x_\ell $
(from a distance).

We postulate that the number of pebbles dropped
is kept in the finite control of the automaton,
so we can unambiguously partition the state set as
 $Q = Q_n \cup \cdots \cup Q_1 \cup Q_0$, 
where $Q_\ell $ consists of states where $\ell $ pebbles are
still available.
The set $Q_n$ contains both initial and accepting states.
Automaton $\cA_\ell $ equals the restriction of the automaton to
the states in $Q_\ell $; we will not specify initial and
accepting states for  $\cA_\ell $, $\ell <n$.

We show how to express the computations of
automaton $\cA_\ell $,
$\ell  \ge 0$,  
on the input tree as FO+DTC$^k$ formulas,
provided we know how to express computations of 
automaton $\cA_{\ell -1}$ if $\ell  \ge 1$.
For  $\cA_\ell $ a matrix $\Phi^{(\ell )}$ is constructed
with predicates $\phi^{(\ell )}_{p,q}$ for $p,q\in Q_\ell $.
These predicates represent the single steps of $\cA_\ell $,
so  
$t \models \phi^{(\ell )\#}_{p,q}(\vec{u},\vec{v})$
iff $\cA_\ell $ has a nonempty computation from
configuration $[p,\vec{u},\alpha]$ to configuration $[q,\vec{v},\alpha]$.
Note that $\Phi^{(\ell )}$ has additional free variables
$x_n, \dots, x_{\ell +1}$ that will hold the positions of the
pebbles already placed on the tree, 
thus representing the pebble stack $\alpha$.

We first study the steps while the pebble $x_\ell$
has not been dropped.
For each of its heads,
automaton $\cA_\ell $  may test 
the presence of one of the pebbles 
$x_n, \dots, x_{\ell +1}$,
or the node label 
or the child number of the 
current node, or 
it may move 
the head up to the parent or down to a specified
child. 
The semantics of these separate instructions, 
relations between the current
and next configurations $[p,\vec{u},\alpha]$ and 
$[q,\vec{v},\alpha]$,
are easily expressed in first-order logic. 
So, we have the following translation table:
\[
\begin{array}{lll}
\mbox{instruction:} & \mbox{formula:}
\\
\tuple{p,\op{up}_{i},q} &
\bigvee_{j} \edg_j(v[i],u[i]) 
\; \land  \bigwedge_{h\neq i} u[h]= v[h]
\\
\tuple{p,\op{down}_{i,j},q} &
\edg_j(u[i],v[i])  
\; \land \bigwedge_{h\neq i} u[h]= v[h]
\\
\tuple{p,\op{lab}_{i,\sigma},q} &
\lab_\sigma(u[i])  
\; \land \bigwedge_{h} u[h]= v[h]
\\
\tuple{p,\op{peb}_{i}(x_m),q} &
u[i] = x_m 
\; \land \bigwedge_{h} u[h]= v[h]
\\
\tuple{p,\op{chno}_{i,j},q} &
(\exists u') \edg_j(u',u[i])
\; \land \bigwedge_{h} u[h]= v[h] & %(j\neq 0)
%\\
%\tuple{p,\op{chno}_{i,0},q} &
%\lnot(\exists u') \bigvee_{j} \edg_j(u',u[i])
%\; \land \bigwedge_{h} u[h]= v[h] & 
\end{array}
\]
or in case of the negative tests
$\negg\op{lab}_{i,\sigma}$,
$\negg\op{peb}_i(x)$, and $\negg\op{chno}_{i,j}$,
the tests above are negated, whereas head positions 
remain unchanged,
e.g., for  $\tuple{p,\negg\op{lab}_{i,\sigma},q}$
the formula is $\lnot\lab_\sigma(u[i]) \land \bigwedge_{h} u[h]= v[h]$.

In general
$\phi^{(\ell)}_{p,q}(\vec{u},\vec{v})$
is a disjunction of such formulas,
as we may have parallel instructions in the automaton.

Additionally when $\ell \ge 1$, 
$\cA_\ell $ may drop pebble $x_\ell $ in state $p$,
simulate $\cA_{\ell -1}$, 
and retrieve pebble $x_\ell $ returning to
state $q$.
Such a `macro step' 
from configuration $[p,\vec{u},\alpha]$ 
to $[q,\vec{v},\alpha]$ is only possible
when there is a pair of pebble
instructions
 $(p,\mbox{drop}_i(x_\ell ),p')$
and  $(q',\mbox{retrieve}(x_\ell),q)$,
such that  $\cA_{\ell -1}$ has a (nonempty) computation from
$[p',\vec{u},\alpha']$ to $[q',\vec{v},\alpha']$, 
with $\alpha' = \alpha (x_\ell,u[i])$.
Hence, 
%for the next `level',  
 $\cA_\ell $ can take a `step' from
$[p,\vec{u},\alpha]$ to $[q,\vec{v},\alpha]$ if the disjunction
of $\phi^{(\ell -1)\#}_{p',q'}(\vec{u},\vec{v})$ 
over all such $q'$ holds,
where the free variable $x_\ell $ in that formula 
is replaced by $u[i]$, the current
position of the $i$-th head of the automaton, 
i.e., the position at which that
pebble is dropped. 
Note that in $\cA_{\ell -1}$, $q'$ has no outgoing instructions
(and hence $q'$ is a final state of $\Phi^{(\ell -1)\#}$). 

Defining the remaining $\phi^{(\ell )}_{p,q}$ to be false,
we obtain a step matrix $\Phi^{(\ell )}$, 
which is deterministic thanks to the determinism 
of the automaton and the semi-determinism of $\Phi^{(\ell-1)\#}$,
cf. Lemma~\ref{Phi}(1).
It is in \mbox{\rm FO+DTC$^k$} by Lemma~\ref{Phi}(3).
The computational behaviour of the automaton $\cA_\ell $ 
is expressed by $\Phi^{(\ell )\#}$, in general,
and more specifically for  $\cA_n$, 
by the disjunction of all formulas 
$\phi^{(n)\#}_{p,q}(\overline{\rt},\overline{\rt})$ 
with $p$ the initial state and $q$ an accepting state.
Note that the last formula is correct by assumption (1)
in the beginning of this proof.
\qed

\noindent
Combining the two inclusions in 
Lemma~\ref{FODTC->DPTWA} and
Lemma~\ref{DPTWA->FODTC},
we immediately get the main result of this paper.
Note that it includes the case of strings.

%% ===============================================
\begin{theorem}
\label{main}
For trees over a ranked alphabet, and $k \ge 1$, 
$\DPTWA{k} = \FODTC{k}$. 
\qed
\end{theorem}

\noindent
As a corollary we may transfer two obvious closure properties
of $\FODTC{k}$,
closure under complement and union,
 to deterministic tree-walking automata with nested
pebbles, where the result is nontrivial.
These properties are a rather direct consequence
of the always-halting normal form in the proof of 
Lemma~\ref{FODTC->DPTWA}, which can be obtained for every 
deterministic automaton.
For $k=1$,
this normal form is further studied with regard to the
number of pebbles needed in \cite{MuSaSe,BSSS}.

%% ===============================================
\begin{corollary} Let $k \ge 1$.
%\leavevmode
%\begin{enumerate}
%\item  $\DPTWA{k}$ is closed under complement and union.
%\item  
   For each deterministic $k$-head tree-walking automaton
   with nested pebbles 
   we can construct an equivalent one that always halts.
%\end{enumerate}
\qed
\end{corollary}

\noindent
When the tree-walking automaton is not deterministic
we no longer can assure the determinism of the formulas
$\Phi^{(\ell)}$ in the proof of Lemma~\ref{DPTWA->FODTC}.
However, by Lemma~\ref{Phi}(2) they are in 
\logic{FO+TC$^{k}$}.

%% ===============================================
\begin{theorem}
\label{Nmain}
For trees over a ranked alphabet, and $k \ge 1$, 
$\NPTWA{k} \subseteq \FOTC{k}$. 
\qed
\end{theorem}

\noindent
The constructions in the proof of Lemma~\ref{DPTWA->FODTC}
use negation in one place only: 
it is used on atomic predicates, to model negative
tests of the automaton (to check there is no specific
pebble on a node). 
Note that negation is not used to construct the formulas 
in the proof of Lemma~\ref{Phi}.
Hence we obtain positive formulas, 
where negation is only used for atomic predicates,
and thus the inclusions $\DPTWA{k} \subseteq \FOposD{k}$
and $\NPTWA{k} \subseteq \FOpos{k}$. 
In the deterministic case negation of a transitive closure 
can (for finite structures) be easily
expressed without the negation \cite{Graedel},
thus in a positive way:  
$\FOposD{k} = \FODTC{k}$. 
With that knowledge the first inclusion is not surprising;
for the nondeterministic case we additionally find a new, positive, 
characterization
(cf. the end of Section~\ref{logpeb}).

\begin{corollary}\label{pos}
For trees over a ranked alphabet, and $k \ge 1$, 
$\NPTWA{k} = \FOpos{k}$.
\qed
\end{corollary}

\noindent
As observed in the Introduction, we do not know whether
$\NPTWA{k}$ is closed under complement
(i.e., whether `\family{pos}' can be dropped from
Corollary~\ref{pos}).
Using the method of \cite{Imm88,Sze},
it is easy to see that, for trees,
$\bigcup_{k\in \natN}\NPTWA{k}$
\emph{is} closed under complement,
which means it is equal to
$\bigcup_{k\in \natN}\FOTC{k}$
(and note that it also equals $\bigcup_{k\in \natN}\NTWA{k}$).

%% ===============================================
%% ===============================================
\section{Single Head on Trees}\label{overview}
%% ===============================================

More than thirty years ago, single-head tree-walking automata 
(with output) were
introduced as a device for syntax-directed translation 
\cite{AhoUll}
(see \cite{EngRozSlu}).
Quite recently they came into fashion again 
as a model for translation of XML specifications
\cite{MilSV,Via,XML,KlaSchSuc,Sch,caterpillars}.

The control of a single-head tree-walking 
automaton is at a single node of the input tree.
Thus it differs %in two aspects 
from the more commonly
known tree automata. These latter automata work either
in a top-down or in a bottom-up fashion and are
inherently parallel in the sense that the control is split
or fused for every branching of the tree.

The power of the classic tree automaton model is well known. 
It accepts the
regular tree languages (both top-down or bottom-up),
although the deterministic top-down variant is less powerful.
For tree-walking automata however, the situation was
unclear for a long time. 
They accept regular tree languages only
\cite{KamSlu,EngRozSlu}, but it
was conjectured in \cite{storage} 
(and later in \cite{trips,walking,caterpillars})
that tree-walking automata cannot 
accept all regular tree languages%
\footnote{Although a footnote in \cite{AhoUll} claims that
the problem was solved by Rabin.}.
This was first proved for `one-visit' automata
(for the deterministic case in \cite{Boj,OkhSalDom},
and for the nondeterministic case in \cite{NevSch}).
Recently  the conjecture was  proved, in a very elegant way, for deterministic 
tree-walking automata in \cite{determinized}, 
and for nondeterministic tree-walking automata in 
\cite{BoCo2}
 (see Examples~\ref{walking} and \ref{DPTWA}).
 
The reason that tree-walking automata cannot
fully evaluate trees like bottom-up tree automata is that
they easily loose their way. When evaluating a subtree it
is in general hard to know when the evaluation has
returned to the root of the subtree. 
In order to facilitate this, in \cite{walking} the 
single-head tree-walking
automaton was equipped with pebbles.
This was motivated by the ability of pebbles 
to help finite-state automata find their way out of
mazes~\cite{BluKoz}.

In  \cite{walking} we have shown that  all
first-order definable tree languages can be accepted by
single-head 
(deterministic) tree-walking automata with nested pebbles,
and that tree languages accepted by single-head (nondeterministic)
tree-walking automata with nested pebbles are all regular.

As observed before, \DLOG\ is the class of languages accepted 
by single-head two-way automata with (nonnested) pebbles
\cite{RitSpr,Pet}.
Thus, for $k=1$ 
(single-head automata vs. unary transitive closure),
our main characterization for tree languages,
Theorem~\ref{main},
can be seen as a `regular' restriction of the result
of Immerman characterizing \DLOG;
on the one hand only 
(single-head) %tree-walking 
automata  with
\emph{nested} pebbles are allowed,
while on the other hand we consider only 
\emph{unary} transitive closure, i.e., transitive closure
for $\phi(x, y)$ where $x,y$ are 
single variables.
Note that unary transitive closure can be simulated in
monadic second-order logic,
which defines the regular tree languages.

\smallskip

We compare the family of tree languages
$\FODTC{1}=\DPTWA{1}$
with several next of kin.
In the diagram below we have five families of languages
\family{xW$^1$A} accepted by 
(single-head) tree-walking automata,
which are either deterministic,  nondeterministic,
or alternating
(\family{D}, \family{N}, or \family{A} in \family{x}),
and may use nested pebbles
in case \family{x} contains \family{P}.
Lines without question mark denote proper inclusion,
those with question mark just inclusion.

%% ===============================================
\begin{figure}
\centerline{%
\unitlength 0.85mm
%\fbox
{\begin{picture}(135,75)(-2,-5)
 \gasset{AHLength=3,Nframe=n,Nadjust=w,Nh=6,Nmr=0,AHnb=0} 
 \node(LFO)(10,00){$\LFO$} 
 \node(DTWA)(40,10){$\DTWA{1}$}
 \node(NTWA)(70,20){$\NTWA{1}$}
 \node(FO)(00,25){$\FO$}
 \node[Nadjust=wh,Nmr=1,Nframe=y](DPTWA)(30,35){$\FODTC{1}=\DPTWA{1}$}
 \node(NPTWA)(60,45){$\FOpos{1} = \NPTWA{1}$}
 \node(TC1)(90,55){\FOTC{1}}
 \node(MSO)(120,65){$\MSO = \REGT$}
 \node(blank)(120,60){$\phantom{\MSO} = \ATWA{1}$}
 \drawedge(LFO,FO){}
 \drawedge[ELside=r](LFO,DTWA){\cite{walking}}
 \drawedge[ELside=r](FO,DPTWA){\cite{walking}}
 \drawedge[ELside=r](DTWA,NTWA){\cite{determinized}}
 \drawedge[ELside=r](DTWA,DPTWA){\cite{determinized}}
 \drawedge[ELpos=40](DPTWA,NPTWA){?}
%
% \drawedge[ELside=r,dash={1.5}0](NTWA,NPTWA){\cite{BoCo2}}
 \drawedge[ELside=r](NTWA,NPTWA){\cite{BoCo2}}
 \drawedge(NPTWA,TC1){?}
 \drawedge(TC1,MSO){?}
\end{picture}}}
\end{figure}

%By $\LFO$ we denote the family of languages definable
%in \emph{local} first-order logic, i.e., 
%dropping the atomic formula $x\le y$.
The inclusion $\LFO \subseteq \DTWA{1}$ was shown in 
\cite{walking}, as well as $\FO \subseteq \DPTWA{1}$.
The regular language $(aa)^*$ cannot
be defined in first-order logic, and shows that
$\DTWA{1} \not\subseteq \FO$.
The strictness of 
$\DTWA{1} \subset \NTWA{1}$  was shown in
\cite{determinized}; their example additionally shows that
$\FO \not\subseteq \DTWA{1}$. 
The result of \cite{BoCo2}
shows even that  $\FO \not\subseteq \NTWA{1}$,
 cf. Example~\ref{walking}.
Logical characterizations of $\DTWA{1}$ and $\NTWA{1}$ 
are given in \cite{NevSch}, also using transitive closure
(but with an additional predicate indicating the 
level of a node modulo some constant).
All families considered here are contained in the
family $\REGT$ of 
regular tree languages that can be characterized by
monadic second-order logic MSO \cite{Don,ThaWri}.
The inclusions 
$\FODTC{1} \subseteq \FOpos{1} \subseteq \FOTC{1}
\subseteq \MSO$ are obvious.
In \cite{Pott} several
logics for regular tree languages are studied;
it is stated as an open problem whether all regular
tree languages can be defined using monadic transitive closure,
i.e., whether
$\FOTC{1} = \MSO$.

Alternating tree-walking automata are
considered in \cite{alternation}.
Alternation combines nondeterminism 
(requiring a successful continuation
from a given state)
with its dual 
(requiring all continuations to be successful).
It is not difficult to see that a (nondeterministic) 
top-down tree automaton can be simulated by an
alternating tree-walking automaton, 
but the reverse inclusion is nontrivial: 
$\REGT= \ATWA{1}$.

If, instead of with pebbles, single-head tree-walking automata
are equipped with a synchronized pushdown or,
equivalently, with `marbles',
then they do recognize all regular tree languages 
\cite{KamSlu,EngRozSlu,trips},
both in the deterministic and nondeterministic case.
Synchronization means that the automaton can push or pop
one symbol when it moves from a parent to a child
or vice versa, respectively.

\para{Questions}
Several of the inclusions between the families of trees
we have
studied are not known to be strict,
cf. the figure in this section.
These are all left as open problems
(but see below).
So, for logics, are the inclusions 
$\FODTC1 \subseteq \FOpos1 \subseteq \FOTC1
\subseteq \MSO$
strict? 
For tree-walking automata,
are the inclusions 
$\DPTWA1 \subseteq  \NPTWA1 \subseteq  \REGT$
strict, 
is $\NTWA1 \subseteq \DPTWA1$?
Considering the use of pebbles,
is there a strict hierarchy for tree languages accepted by
(deterministic) tree-walking automata in the number of
pebbles these automata use? 

The pebbles we have used in this paper are considered as
pointers, and differ from the pebbles that can only be picked up
at the node they are dropped. We conjecture that our
type of pebbles is more powerful than the usual one.
For nonnested pebbles the two types have the same power,
even when the number of pebbles is fixed
(as shown in Theorem~2.2 of \cite{BluHew}).

Alternating tree-walking automata with classic 
(nested) pebbles were considered
in \cite{MilSV}, where it is shown that these automata accept
the regular tree languages. We did not investigate whether
alternation together with pointer pebbles again yields the
regular languages,
i.e., whether $\APTWA{1} = \REGT$.

\para{Answers}
Since the appearance of the report version of this paper,
several of the questions above have been answered.

In the recent \cite{BSSS} it is shown that, surprisingly, 
the two types of nested pebbles have the same power. Moreover, it is 
shown that the number of pebbles gives rise to a strict hierarchy, both 
for deterministic and nondeterministic automata, and that the inclusion
$\NPTWA1 \subset \REGT$ is strict.

In the very recent paper \cite{Muz06} it is shown that 
$\APTWA{1} = \REGT$ holds, using an easy variation of the proof 
technique of \cite{MilSV}.

%\para{Trips}
Both tree-walking automata and logical formulas can also be used
in a natural way to define binary relations on the nodes of trees,
called \emph{trips} in \cite{trips}.
It is straightforward to show that Theorem~\ref{main} and 
Corollary~\ref{pos} also hold for trips.
XPath-like formalisms (relevant to XML) that are equivalent with,
or closely related to, the trips defined in 
\logic{FO+posTC$^1$}
and
\logic{FO+TC$^1$} 
have recently been studied in \cite{GorMar05,Cat06}.

\para{Single head on strings}
For strings the equivalence $\MSO = \REGT$ between 
monadic second-order logic and finite-state automata
was obtained independently by 
B\"uchi, Elgot, and 
Trakhtenbrot
\cite{Buchi,Elg61,Tra62}.
As observed, tree-walking automata on monadic trees correspond to
two-way finite-state automata on strings. 
Since the latter are equivalent to
ordinary finite-state automata (both deterministic and 
nondeterministic) \cite{RabSco,She} most of the hierarchy 
is known to collapse in the string case: $\DTWA{1} = \REGT$.
Hence for $k=1$ our main result gives two additional
characterizations of the regular string languages
$\REGT = \NPTWA{1}$ and $\REGT = \FODTC{1}$.
The second equality was shown in \cite{BarMak}
 (see also Exercise~8.6.3 of \cite{EbbFlu},
 and Satz~2.0.1 of \cite{Pott}).

%% ===============================================
%% ===============================================
\section{Walking on Graphs}\label{graphs}
%% ===============================================

We generalize our results on trees 
(and strings) to more general
families of graphs. There is a basic problem that
has to be resolved.
We have to propose a model of graphs that is
suitable for both graph-walking automata and logic. 
The structures for which Immerman has obtained his
logical characterization of \DLOG\ are ordered, i.e.,
they are equipped with a total order $\le$, which can either be
specified directly or, as transitive closure is
available in the logical language, by a direct successor
relation. 
In fact, when placing a structure on the tape of a
Turing Machine, as has to be done for \DLOG, it almost
automatically obtains an implicit order on its elements.
A natural way to present a structure 
(assuming only relations of at most arity two)
as input
to a graph-walking automaton is to represent all pairs
in a binary relation as directed edges in a graph,
labelled by a symbol representing the relation;
unary relations and constants are translated as labels
for the nodes.
If we fully represent a total order $\le$ in this way,
there will be many edges with the same label leaving (and entering)
each node. Although the edge to the direct successor can
be directly specified in first-order logic 
we see no natural machine model
that can choose this edge among the many candidates.

A solution would be to assume the existence of
direct successor edges in the graph, 
i.e., to consider `ordered' graphs
(cf. the end of the Introduction),
but we feel this 
requirement is too restrictive.
On trees, tree-walking automata use 
the preorder to traverse the nodes of
the input. Thus, the order is implicit in the tree,
and not explicitly given by edges with a special label.
%A graph-walking automaton has no explicit storage from
%which an ordering can be inferred, so such an ordering must
%be obtained in another fashion. 
Below we generalize this idea to graphs: we do not assume
an explicit total order on the nodes, but we postulate
the existence of a special graph-walking automaton, 
the \emph{guide}, that can be used as
a `subroutine' to traverse
the nodes of any graph in the family of graphs we are
considering. For trees the guide follows the preorder,
for (two-dimensional) grids the guide visits the nodes
row by row (with generalizations to higher dimensions).

%% ===============================================
\para{Graphs}
We consider (families of) nonempty 
finite directed graphs, where nodes and edges
are labelled using a common alphabet $\Sigma$. 
In order to be able to locally distinguish edges we
require that no node has two outgoing edges or two
incoming edges  with the same label.
Hence the graphs we consider are of bounded degree.
Parallel edges and loops are allowed.
Thus, each edge label represents a (partial) injective function
on the set of nodes.
We consider (weakly) connected graphs only.

Trees over a ranked alphabet fall under this definition
since we label the edge from a parent to its $i$-th child by $i$.
Another example of such a family is formed by grids, where
two different labels are used to distinguish edges to
horizontal and vertical neighbours. (In our model there
is no need to introduce special node labels that mark the
boundaries of the grid, cf. \cite{BluHew,BarMak},
 as the labels of the edges incident to a node 
are available to our automata). 

%% ===============================================
\para{Automata}
A $k$-head 
\emph{graph-walking automaton with nested pebbles}
is like its relative for trees in that it visits the
nodes of a graph, 
each of its heads walking along the edges from node to
node. 
It may check the labels of its current nodes, 
check whether such a node has an incident incoming/outgoing 
edge with a specific label
(generalizing the concept of child number
and the rankedness of $\Sigma$, respectively), 
check for the presence of a specific pebble
(from a finite set), 
and it may then move each head along any edge in either
direction choosing the edge based on its label.
Moreover,  
it may drop and retrieve pebbles in a nested fashion, 
as before.

Formally we change the operations and tests available to
the automaton.
The moves $\op{up}_i$ and $\op{down}_{i,j}$
are replaced by the more symmetric pair
$\op{inmove}_{i,\sigma}$, $\op{outmove}_{i,\sigma}$
to specify a move by the $i$-th head 
along an incoming (outgoing) edge with label $\sigma$.
Likewise we replace the child-number test
$\op{chno}_{i,j}$
by the tests
$\op{inedge}_{i,\sigma}$ and $\op{outedge}_{i,\sigma}$
on the existence of an 
incoming (outgoing) edge with label $\sigma$ for the
node currently under the $i$-th head.

\smallskip
Generally graphs do not have a distinguished node
(like the root is a distinguished node for trees)
and we change the definition of acceptance accordingly.
A graph-walking automaton  with nested pebbles
accepts a given input graph if
the automaton has an accepting computation
(starting in the initial state, halting in an accepting
state) when started with all its heads on
 \emph{any} node of the input graph;
initially and finally there should be no pebbles on the
graph.
We require that the existence or nonexistence
of an accepting computation
is independent of the chosen initial node,
as this seems a natural condition, 
especially in the context of determinism.
Note that not all automata satisfy this requirement.%
\footnote{%
Formally, we define $L(\cA)$ to consist of all graphs $g$ over
$\Sigma$ such that $\mbox{acc}(u)$
for every node $u$ of $g$, where $\mbox{acc}(u)$
means that
$[q_0,u^k,\varepsilon] \vdash^*_{\cA,g} [q,\vec{v},\varepsilon]$
%for some $q\in A$ and some $k$-tuple $\vec{v}$ of nodes of $g$
for some halting configuration $[q,\vec{v},\varepsilon]$ with
    $q \in A$ and $\vec{v}$ is a $k$-tuple of nodes of $g$
(and $u^k$ is the $k$-tuple of $k$ copies of $u$).
Alternatively, we could require this for \emph{some} node
$u$ of $g$, because, as discussed above, $\cA$ is restricted
to satisfy the requirement that for all graphs $g$ over $\Sigma$
and all nodes $u_1,u_2$ of $g$,
if $\mbox{acc}(u_1)$ then $\mbox{acc}(u_2)$.
We note, however, that this restriction is not essential for
our results: it is easily shown that for searchable graphs
(see below) the restriction can be dropped.%
}

For trees over a ranked alphabet this definition is
obviously equivalent to the old one as a
tree-walking automaton 
can always move all its heads
to the root.
Disregarding pebbles,
for grids the definition is equivalent to the $k$-head
automaton of \cite{BarMak}
 (and for $k=1$ to the 2-dimensional automaton of \cite{BluHew}).

In order to avoid an abundance of new notation,
we keep the notation \DPTWA{k}, etc., 
for families accepted by (deterministic) $k$-head automata 
(with nested pebbles)
even in the more general context of graphs.

%% ===============================================
\para{Logic}
The first-order logic for graphs over the label alphabet
$\Sigma$ has atomic formulas
$\lab_\sigma(x)$, 
 $\sigma \in \Sigma$, for a node $x$ with label $\sigma$,
$\edg_\sigma(x,y)$, $\sigma \in \Sigma$, 
for an edge from $x$ to $y$ with label $\sigma$,
and
$x=y$.

As for automata, we keep the notation for families
defined by our first-order logic with several variants of 
transitive closure.
Note that we do not allow the predicate $x\le y$,
which makes the logic more like the local variant 
of first-order logic. 
As we have seen in Section~\ref{sec:prel},
for trees this predicate is definable in first-order logic with
(positive deterministic) transitive closure,
and the families $\FODTC{k}$, etc., 
of tree languages definable in first-order logic with
transitive closure,
do not change by this restriction.

\smallskip
For arbitrary families of graphs the computation of
a graph-walking automaton with nested pebbles can be specified
in first-order logic with transitive closure, 
like in Section~\ref{peblog}.

%% ===============================================
\begin{lemma}\label{graphs:DPTWA->FODTC}
For every family of graphs, and $k \ge 1$, 
$\DPTWA{k} \subseteq \FODTC{k}$.
\end{lemma}
\proof
In the proof of Lemma~\ref{DPTWA->FODTC}, 
where we consider trees
instead of graphs,
no reference is made to properties particular to that domain.
The only place where the proof given has to be adapted is where
the accepting condition is phrased in the logic. 
Here any node
must have an accepting computation when we start the
automaton with all heads on that node, 
so we write 
$(\forall x)(\exists \vec{y})\bigvee_{q \in A}
\phi_{p,q}^{(n)\#}(x^k,\vec{y})$ 
where $p$ is the initial state,
the disjunction  is taken over 
the set $A$ of accepting states,
and $x^k$ is the $k$-tuple consisting of $k$ copies of $x$.
Note that we might as well take 
$\exists x$ instead of $\forall x$.
\qed

\noindent
The reverse inclusion needs an additional notion,
the ability to search each graph in a given family of graphs.

%% ===============================================
\para{Searchable Graphs}
A family of graphs is \emph{searchable}
if there exists a (fixed) 
single-head
deterministic graph-walking 
automaton with nested pebbles 
that, for each graph in the family,
and each node of the graph, when started in that node
in the initial state
the automaton halts after completing a walk along the 
graph during which each node is visited at least once.  
Pebbles may be used in a nested fashion during the
walk, as before.
Thus the automaton serves as a \emph{guide} for the family
of graphs, and makes it possible to 
establish and traverse a total order of 
the nodes of the graph,
generalizing the concept of preorder traversal used for
trees.
(In fact, the guide visits the same node perhaps several times,
like in a preorder traversal on trees, but we will see that
the unique first visit to a node can be recognized.)
Note that the total order may depend on the node 
at which the guide starts its walk.
Note also that if a family of graphs is searchable, then any
subset of the family is searchable (by the same guide).

\smallskip
We give several elementary examples of this notion.
Trees over a ranked alphabet form a searchable family:
from any node we may walk to the root to start a preorder
traversal through the tree.
Similarly binary trees --which are basically ranked trees where
every node has rank two, but where either child of a node can
be missing-- are a searchable graph family.

Unranked (ordered) trees, where there is no bound on the
number of children of a node, are important in the
theory of data representation using XML \cite{XML,KlaSchSuc,Sch}.
Here they are considered in their natural encoding as
binary trees, where each node carries 
possibly two pointers (edges),
one to its first child, one to its right sibling.
In this way they are  a searchable graph family.
The (single-head) automata we obtain using this representation
may move to the first child or to the next sibling of a node
(and back),
exactly as customary in the 
literature \cite{XML,OkhSalDom} 
(albeit without pebbles).
For an overview on logics for unranked trees, see \cite{Lib06}.

Rectangular (directed) grids, 
edges pointing to the right or downwards, 
with edge labels distinguishing these two types of
edges, form another example.
This can be generalized to higher dimensional grids%
\footnote{%
A $d$-dimensional grid has nodes $(x_1,\dots,x_d)$
with $x_i \in \natN$, 
$1\le x_i \le k_i$, for certain $k_1, \dots,k_d$.
For each $i$ it has $i$-labelled edges from 
$(x_1,\dots,x_i,\dots,x_d)$ to
$(x_1,\dots,x_i +1,\dots,x_d)$, provided $x_i < k_i$.
}
\cite{BarMak}.

All the above are superseded by
acyclic, connected graphs with a single source
(and the local restriction on edge labels).
We can effectively turn those graphs into trees by
placing an ordering on the labels and ignoring
all incoming edges of a node except the edge which
has minimal label among those edges.
Graph-walking automata (with one head and no pebbles)
on such graphs were considered in \cite{KamSlu}.

Obviously, a family of graphs such that, 
for some $\Sigma'\subseteq \Sigma$,
the graphs induced by the $\Sigma'$-labelled edges
belong to one of the above families, is searchable too.
This includes the family of all `ordered' graphs.

\smallskip
The above families are traversed in a rather standard way.
In the next example we use a pebble to find and `break' a
cycle.

%% ===============================================
\begin{example}
Consider an ordinary binary tree, and take one of its
leaves. If we identify this leaf with the root
(i.e., we remove it and redirect the incoming edge to the root)
we obtain a single directed cycle from which trees radiate,
like charms hanging from a bracelet. 

We argue that the family of these graphs is searchable. If
started on a node in the cycle a graph-walking automaton
may drop a pebble at that node. This effectively reduces
the graph to a tree with the marked node as root, and a
preorder traversal can be made.

We claim that a node on the cycle can be found,
with the help of a single pebble, starting on an arbitrary node 
of the graph. Put the pebble on the initial node.
Search the graph `below' the pebble as if it is a tree;
if indeed it is a tree, then we finally return to the pebble from
below. Otherwise we are on the cycle, 
and we enter the leaf that has been identified
with the root, and will find the pebble from above.
Thus, in order to find a node on the cycle we repeat 
the above search, each time
moving the pebble one node up while not on the cycle.
\qed
\end{example}

\noindent
Cyclic grids, or toruses, where the last node of each
row has an edge to the first node of that row, 
and similarly for columns,
can be searched using two pebbles.
We search the grid row-by-row: 
the first pebble marks the position we start with
(in order to stop when all rows are visited; we do not move
this pebble during the traversal), 
the second pebble moves down in the first column to mark  
the position
in which we started the row (in order to stop when we
finish the row; we then move the pebble down to the next row
until we meet the first pebble). 
This can be generalized to $d$ pebbles in the $d$-dimensional
case.%
\footnote{%
A $d$-dimensional torus is a $d$-dimensional grid
with additional $i$-labelled edges from
$(x_1,\dots,k_i,\dots,x_d)$ to
$(x_1,\dots,1,\dots,x_d)$. 
}

It is an open question whether mazes
(modelled as connected subgraphs of grids) form a 
searchable family of graphs, that is, with a single head and the 
help of nested pebbles.
They cannot be searched without the help of pebbles,
or with only a single (classic) pebble
\cite{Bud,Hof}.
According to \cite{BluKoz} these graphs can be searched
by single-head graph-walking automata with two (classic) pebbles
which are unfortunately not used in a nested fashion,
or using two heads (without pebbles).
%Ilcinkas slides : 
%Rollik, Acta Inf, 1980, Automaten in planaren Graphen
%no finite team of automata
%(local cooperation)
%global => hydra
%Fraigniaud, Ilcinkas, Markou, Pelc
%2-head 'hydra' not less finite team
%k-head hyda more than team k local
%3-head more than 2-head hydra (open: in general?)
%3-head hydra not universal
%conjecture: no hydra universal

The family of all graphs 
(over a given alphabet with at least two elements)
is not searchable, not even with nonnested pebbles
or with several heads.
This follows from results of Cook and Rackoff \cite{CooRac}.
The deterministic $k$-head graph-walking automaton with
$n$ pebbles is a special case of the jumping automaton
of \cite{CooRac} with $k+n$ pebbles.
A jumping automaton may move its pebbles along the edges of
the graph, or move one of its pebbles to the location of 
another pebble (`jumping').
Retrievability of pebbles at a distance is a particular
case of this jumping facility.
Basically, \cite[Theorem~4.9]{CooRac} states that 
the number of pebbles needed to visit all nodes of a
$d$-dimensional torus grows as a function of $d$. 
Then \cite[Theorem~4.13]{CooRac}  concludes there is
no jumping automaton that searches all graphs over a three letter alphabet,
by coding arbitrary graph alphabets 
(like the $d$ dimension edge labels for toruses)
into a three letter alphabet. 
A slight adaptation of the proof (because of the graph
model used) makes it valid in our setting, with two letters instead of three.
For classic pebbles (and a single head) it is shown in \cite{Rol} that the 
family of all planar graphs is not searchable.

For an overview of approaches to the exploration of mazes and
graphs by finite automata see \cite{Frai+04,BBRRT}.

\smallskip
For searchable graphs one can obtain the converse inclusion
from our previous result, and translate logic into automata.

%% ===============================================
\begin{theorem}\label{searchable}
For every searchable family of graphs, and $k \ge 1$, 
$\DPTWA{k} = \FODTC{k}$.
\end{theorem}

\proof
By the previous lemma it suffices to show 
$\FODTC{k} \subseteq \DPTWA{k}$
for searchable graphs, i.e., to reconsider the proof of 
Lemma~\ref{FODTC->DPTWA}.
There is no need to return to a `root' 
(we do not have one)
to test acceptance
by some automaton 
(the induction hypothesis used as a `subroutine');
in the definition of acceptance we have 
required acceptance for any initial node, so we merely move all 
heads to the same position: drop a pebble, let each
head search for it, and lift the pebble.%
\footnote{%
Dropping a temporary pebble is also a technique to show 
we may assume
heads to be sensing, i.e., the automaton
 can check whether two heads occupy
the same position.}

% search the tree for a node => straigthforward guide.
% check all nodes, existence node (also OK)
% Sipser: move head to successor may see same node several
% times also OK !?

At several points in that proof the automaton is required to 
make a systematic traversal through the input tree,
or more specifically, to find the preorder successor of a
given node, a function easily implemented for trees.
At some other points we are supposed to move a head to
a specific position marked by a pebble: the pebble again is found 
by a systematic traversal of the tree.

Here we instead use the `guide', 
the automaton that makes the family searchable, 
as a subroutine to determine the successor of a node 
(and to make traversals of the graph).
We have to mind some details, however.
First, the guide may use its own pebbles during its computation,
as is required to traverse a torus;
leaving these pebbles on the graph would block the removal
of the other pebbles on the graph.
Second, 
during the traversal of the graph the guide may visit 
the same node several times. 
In order to define 
the notion of successor properly, we
have to distinguish a particular visit;
for this we take
the first visit to the node
(as in the preorder of the nodes of a tree).
Third, the order of visiting the nodes 
may depend on the node 
at which the guide is started, which makes the
notion of successor undefined (think torus again).

Thus, we need to explain how the unique successor can be 
found for each node in the graph (except the last). 
First, at the very beginning of its computation, the automaton we
construct for a given formula drops a pebble on the node
where it is started (i.e., the node where all its heads 
are initially positioned; this may be any node of the graph).
This pebble, let us call it `the origin', remains at its position
during the full computation to serve as a fixed position in the 
graph. 
Every successor will be determined in the order when
starting the guide in the origin. Now the task of finding 
the successor of a (marked)
node using a single head is executed as 
follows. First we use the guide 
(from wherever we are) to move the head to the origin. 
When the origin is reached, we `reset' the guide
(removing its pebbles from the graph, 
returning to its initial state)
and again start a traversal of the graph looking for our
marked node. 
When we reach the marked node,
we continue running the guide until we reach a node that
is visited for the first time: our (well-defined) successor.
In order to check whether a node is visited for the first
time we leave all pebbles of the guide at their
position and drop a new pebble on the node. 
We then start a copy of the guide
at the origin and run it until we find the new pebble.
If the copy of the guide is in the same state and
has the same positions for its pebbles as
the original guide, then it visits the node for the first time:
due to its determinism the guide cannot visit a node twice in the
same configuration. After this test we stop the copy of the guide
and retrieve its pebbles. We either have found the successor node,
retrieve the pebbles of the guide and proceed as needed,
or we lift the new pebble and continue the guide.
If the guide halts without finding a first visit, 
the marked node was the last node in the order.
\qed

\noindent
As in Corollary~\ref{pos}, 
we have in the nondeterministic case
$\NPTWA{k} = \FOpos{k}$
for every searchable family of graphs, and $k \ge 1$.

\para{Multi-Searchable Graphs}
It is open whether we can search a maze 
(a connected subgraph of a grid)
with a single head using nested pebbles. 
However with two heads we can search a maze \cite{BluKoz}. 
To cover this family we need to extend
the notion of searchability:
as the automata we are dealing with have $k$ heads,
it seems only fair to extend the guide with this commodity.
A family of graphs is \emph{$k$-searchable} if there is a 
deterministic guide
as before, which now may have $k$ heads.
The guide starts its computation with all
its heads on an arbitrary node of the graph,
and performs a traversal of the graph.
Recall that the formal model assumes
that at most one head moves in each computation step
of the guide; this ensures that at most one node is visited
for the first time, thus uniquely defining the order of the nodes.

With this notion the results of the previous section 
can be extended to the larger class of 
$k$-searchable graphs.
But there is a catch: we have to extend our automaton
model with a \emph{new instruction} 
$\tuple{p,\op{jump}_i(x),q}$
that moves a given
head $i$ to the position of a given pebble $x$ 
(like the `jumping' instruction from \cite{CooRac}). 
In the case of searchable graphs a single head 
can be moved to any pebble, just by running the guide 
using that head, searching for the pebble.
For $k>1$ however, 
in order to move a single head to a pebble,
we need several heads, either introducing
auxiliary heads (loosing the connection between 
number of heads
and arity of transitive closure) or loosing the 
position originally held by the heads needed for the search.
(Even ignoring the problem of 
how to move all heads to the same initial
position as needed for the guide.)

The suggested additional jump-instruction is rather natural if we
think of pebbles as pointers as we did before.
If the graph-walking automaton stores a finite number
of `addresses' of nodes, it can use a simple assignment
to move a head to such a position.
Adding this instruction is `backward compatible': 
it will not change any of the previous results.
On families of searchable graphs (which includes trees)
the new instruction can be implemented as explained above.

Note that if a family is $k'$-searchable, then it is
$k$-searchable for $k \ge k'$.
Thus, the next result generalizes Theorem~\ref{searchable}
(which for trees is Theorem~\ref{main}).

\begin{theorem}\label{k-searchable}
Let the automaton model be extended with the jump-instructions,
as discussed above. 
For every $k$-searchable family of graphs, $k \ge 1$, 
$\DPTWA{k} = \FODTC{k}$.
\end{theorem}
\proof
For the inclusion $\DPTWA{k} \subseteq \FODTC{k}$
we note that the new instruction (move head $i$ to pebble
$x$) is directly expressible in the logic as we assume variables
for the positions of heads and pebbles, 
cf. the proof of Lemma~\ref{DPTWA->FODTC}.

For the converse inclusion $\FODTC{k} \subseteq \DPTWA{k}$
we once more carefully
inspect the proof of Lemma~\ref{FODTC->DPTWA},
or rather the proof of Theorem~\ref{searchable}.

As observed in the latter proof,
the guide is used for two purposes:
(1)~to find the successor of a node, and
(2)~to move a specific head to a specific pebble
(and note that (2) is also used in (1):
a kind of bootstrapping).
For (2) we now use the new instruction, whereas for 
(1) we use the $k$-head guide.
Note that to initiate the guide, all heads can be moved to
the origin by the new instruction.
Note also that before initiating the copy of the guide, 
pebbles should be dropped on all current head positions of the guide
(to be able to compare them with those of the copy of the guide, 
and to 
restore them in case the successor has not been found).
\qed

\noindent
Since mazes are $2$-searchable, we obtain that for mazes
and $k \ge 2$, $\DPTWA{k} = \FODTC{k}$.

Again we obtain a similar result for the nondeterministic case:
$\NPTWA{k} = \FOposD{k}$
for every $k$-searchable family of graphs.

%It seems unknown whether the set of all graphs is 
%$k$-searchable for some $k$, even without pebbles.
As discussed before Theorem~\ref{searchable}, our 
$k$-head graph-walking automaton with nested 
pebbles is a special case of the jumping automaton
of \cite{CooRac}; this includes the `jumping' instruction 
of moving a head to the position of a pebble.
Thus, by the results of \cite{CooRac}, 
the family of all graphs (over $\Sigma\supseteq\{0,1\}$) is not 
$k$-searchable for any $k$.

\smallskip
According to Theorem~\ref{k-searchable},
the existence of a $k$-head guide for a family of graphs entails that 
every language in $\FODTC{k}$ can be implemented by a
$k$-head graph-walking automaton with nested pebbles.
This is obvious for the language $L_0$ defined by the formula
$(\forall x) \lab_0(x)$: 
it can be checked directly by the guide
whether every node visited has label $0$.
It is easy to see that this also works the other way
around:
assuming $\Sigma\supseteq\{0,1\}$,
an automaton for $L_0$ must visit all nodes of 
any input graph that has only node label $0$, 
since otherwise it also accepts that graph with the labels
of unvisited nodes changed into $1$.
The automaton 
can be turned into a guide for the family by 
making it behave as if every node label equals $0$.
Here we require that the family is
\emph{node-label-insensitive}: 
membership of a graph in the family 
does not depend on the label of any node of the graph.
For instance, the families of unranked trees, 
grids, and toruses we considered
are node-label-insensitive.
%\footnote{Strictly speaking, the family of trees over a
%ranked alphabet is not node-label-insensitive, as the node
%label depends on the arity of the node. We can include ranked
%trees by demanding the property for each rank separately.}
Thus, a node-label-insensitive family of graphs is
$k$-searchable iff $L_0 \in \DPTWA{k}$.
Hence for such a family Theorem~\ref{k-searchable}
also holds in the other direction:
if $\mathcal F$ is a node-label-insensitive family of graphs, 
then for every $k$, 
$\mathcal F$ is $k$-searchable iff
$\DPTWA{k} = \FODTC{k}$.
Moreover, 
$\mathcal F$ is $k$-searchable for some $k$ iff
$\DPTWA{} = \FODTC{}$,
where
%$\DPTWA{}$ and $\FODTC{}$ are like
%$\DPTWA{k}$ and $\FODTC{k}$, without restriction on $k$.
$\DPTWA{} = \bigcup_{k\in \natN}\DPTWA{k}$, and
%  = \bigcup_{k\in \natN}\DTWA{k}$, and
$\FODTC{}= \bigcup_{k\in \natN}\FODTC{k}$.
Moreover, this holds iff 
$\bigcup_{k\in \natN}\DTWA{k} = \FODTC{}$,
assuming that the $k$-head graph-walking automaton
(without pebbles) has an additional instruction to move
one head to another 
(i.e., it is the jumping automaton of \cite{CooRac}).

Applying this to the family of all graphs,
which is clearly node-label-insensitive,
we note that the inclusions
$\DPTWA{k} \subset \FODTC{k}$ 
(cf. Lemma~\ref{graphs:DPTWA->FODTC}) and
$\DPTWA{} \subset \FODTC{}$ are strict,
the language $L_0$ being in the difference.

\para{Pebbles left unturned}
An obvious question that remains open is whether 
there exists a strict hierarchy in \DLOG\ with respect
to the number of heads used (allowing nested pebbles),
or equivalently with respect to the arity of transitive
closure.
The same question can be asked for trees and grids.
For arbitrary graphs, strictness of the hierarchies 
$\FOTC{k}$ and $\FODTC{k}$
is shown in \cite{Gro}.
Some other unresolved questions were stated in 
Section~\ref{overview}, regarding single-head automata on trees.
The question concerning the nature of our nested pebbles,
whether the pointer model is more powerful than the classic
model, remains interesting for multi-head automata.
Another question is whether our results can be generalized
to alternating automata and the alternating transitive closure
operator of \cite{Imm87}.

%% ===============================================
\section*{Acknowledgements}
We thank Frank Neven and Thomas Schwentick, the authors of \cite{NevSch}
 (a first version appeared at ICALP 2000) 
where our result $\FODTC{1}=\DPTWA{1}$ is cited,
for their patience.
We also thank them, Luc Segoufin, and 
David Ilcinkas for many useful comments.
Finally, we thank the referees for their constructive remarks.

\bibliographystyle{alpha}
%% ===============================================

\end{document}